\documentclass[aps,prd,superscriptaddress,showpacs,reprint]{revtex4-1}
\usepackage[utf8]{inputenc}
\usepackage{textcomp}
\usepackage[dvipsnames]{xcolor}
\usepackage{amssymb,amsfonts,amsmath}
\usepackage[sort&compress]{natbib}
\usepackage[english]{babel}
\usepackage{braket}
\usepackage{graphicx}
\usepackage{bm}
\usepackage{bbm}
\usepackage{braket}
\usepackage{IEEEtrantools}
\usepackage{pstricks,pst-coil,pst-node,pst-circ,pst-plot,pst-3dplot,
pst-solides3d,pst-sigsys,pstricks-add,pst-eucl}
\usepackage[normalem]{ulem}
\allowdisplaybreaks[1]
\def\biggg#1{{\hbox{$\left#1\vbox to30.0pt{}\right.$}}}

\def\Biggg#1{{\hbox{$\left#1\vbox to50.0pt{}\right.$}}}

\DeclareMathOperator{\tr}{tr}
\usepackage[caption=false]{subfig}

\begin{document}

\title{Dynamical generation of low-energy couplings
from quark-meson fluctuations}

\author{Florian Divotgey}
\email[]{fdivotgey@th.physik.uni-frankfurt.de}
\affiliation{Institut f\"ur Theoretische Physik, 
Johann Wolfgang Goethe-Universit\"at, 
Max-von-Laue-Stra\ss e 1, 
D-60438 Frankfurt am Main, Germany}

\author{J\"urgen Eser}
\email[]{eser@th.physik.uni-frankfurt.de}
\affiliation{Institut f\"ur Theoretische Physik, 
Johann Wolfgang Goethe-Universit\"at, 
Max-von-Laue-Stra\ss e 1, 
D-60438 Frankfurt am Main, Germany}

\author{Mario Mitter}
\email[]{mitter@bnl.gov}
\affiliation{Department of Physics, Brookhaven National Laboratory, Upton, NY 11973, USA}

\date{\today}

\begin{abstract}
We extend our recent computation of the low-energy limit of the linear 
$O(4)$ Quark-Meson Model. The present analysis focuses on the 
transformation of the resulting effective action into a nonlinearly realized 
effective pion action, whose higher-derivative interaction terms are parametrized 
by so-called low-energy couplings. Their counterparts in the linear model are determined 
from the Functional Renormalization Group flow of the momentum-dependent four-pion vertex, 
which is calculated in a fully $O(4)$-symmetric approximation by including also momentum-dependent 
$\sigma\pi$ interactions as well as $\sigma$ self-interactions. Consequently, these higher-derivative 
couplings are dynamically generated solely from quark and meson fluctuations, initialized 
at a hadronic scale. Despite our restriction to low-energy degrees of freedom, we find that 
the qualitative features of the fluctuation dynamics allow us to comment on the range of 
validity and on appropriate renormalization scales for purely pionic effective models.
\end{abstract}

\maketitle

\section{Introduction}
\label{sec:introduction}

A time-honored possibility to study the low-energy regime of the theory of strong interactions, 
Quan\-tum Chromo\-dy\-na\-mics (QCD), is given by Effective Field Theories (EFTs). 
In this approach, one investigates QCD-inspired effective models that describe the interactions 
of the relevant low-energy degrees of freedom, namely, those of hadrons. In order to capture the 
low-energy dynamics of the strong interaction properly, the construction of these models is 
mainly based on the internal and spacetime symmetries of QCD and their possible breaking. 

In the context of low-energy models for QCD, the most central symmetry is given by the 
chiral $SU(N_{f})_{L}\times SU(N_{f})_{R}$ symmetry accidentally arising from the quark 
sector of the QCD Lagrangian. Here, $N_{f}$ refers to the number of dynamical quark flavors, 
which will be set to two, $N_{f}=2$, throughout the rest of this work. 

The importance of chiral symmetry is twofold: On the one hand, the hadronic 
currents arise as multiplets with respect to the chiral group. Therefore, group theory allows 
for a systematic construction of chirally invariant Lagrangians. On the other hand, the explicit 
and spontaneous breaking of chiral symmetry is of immediate physical relevance. 
In particular, one observes so-called pseudo 
Nambu-Goldstone bosons (pNGBs) in the associated particle spectrum. Since these particles are 
very light, they dominate the low-energy dynamics of the theory and, consequently, are
of crucial importance for a proper low-energy description. In the case of two-flavor QCD, 
these pNGBs are associated with the pion fields. 

In the framework of QCD, the most important example of an EFT is given by Chiral 
Perturbation Theory (ChPT) \cite{Gasser:1983yg, Gasser:1984gg}. Conceptually, this approach 
corresponds to a simultaneous expansion of the QCD generating functional in powers of 
pion momenta and quark masses. The associated effective Lagrangian contains an infinite tower 
of pion self-interactions coupled by so-called low-energy constants (LECs). 
These coupling constants encode essential information about the low-energy 
regime of QCD. At lowest order in the chiral expansion, the ChPT Lagrangian is equivalent to the
Nonlinear Sigma Model (NLSM) \cite{GellMann:1960np}.

Apart from ChPT, it is also possible to construct effective low-energy models 
from a linear realization of chiral symmetry. The resulting family of models is usually referred 
to as Linear Sigma Models (LSMs). In contrast to the nonlinear models, the pNGB fields and their 
chiral partners are here treated on the same footing. The simplest and most 
prominent example of such a model is given by the $O(4)$ LSM, which describes the interaction of the 
$\sigma$ meson and the three pions.

In a recent work \cite{Eser:2018jqo}, we studied the low-energy 
limit of the $O(4)$ LSM coupled to quarks, the so-called Quark-Meson Model (QMM), within 
the Functional Renormalization Group (FRG) approach. Besides
the Yukawa coupling of the scalar and pseudoscalar mesons to the quark fields, 
the $O(4)$ LSM has been extended by complete sets of derivative couplings of 
order $\mathcal{O}(\partial^{2})$ and $\mathcal{O}(\partial^{4})$ 
for the pion fields. This corresponds to an approximation well beyond 
the usual local potential approximation (LPA) and LPA' truncations, where these 
derivative couplings are solely generated from meson and 
quark fluctuations. 

After its calculation from the FRG flow, the effective action of the QMM has 
been reduced to an effective pion action by integrating out the $\sigma$ field, 
similarly to Refs.\ \cite{Jungnickel:1997yu, Divotgey:2016pst}. In this action, the 
higher-derivative terms are parametrized by the low-energy couplings of the $O(4)$ QMM.

In the present work, we improve and extend
the previous analysis of these low-energy couplings in several crucial ways: 
\begin{itemize}
    \item [(i)] We transform the effective action of the QMM into a nonlinearly
    realized pion action by explicitly restricting the dynamics of the system onto the vacuum manifold 
    $SO(4)/SO(3)$. This manifold is associated to the spontaneous breakdown of the $O(4)$ symmetry.
    After choosing a specific set of coordinates on $SO(4)/SO(3)$, we will deduce relations between the 
    higher-derivative couplings in      
    the linearly realized QMM and the nonlinear model featuring only pion dynamics. This nonlinear model is
    then referred to as the low-energy limit of the QMM within this work.
    \item [(ii)] In order to determine the range of validity of the low-energy effective theory, we    
    investigate the relative importance of mesonic and fermionic loop contributions to the renormalization   
    group (RG)-scale dependence of the low-energy couplings in the nonlinear model.
	\item [(iii)] We introduce the higher-derivative couplings in the linear model 
	in a completely $O(4)$-invariant 
    manner, i.e., taking also momentum dependences of the interaction of the pion fields with the
    $\sigma$ meson into account, and also include a scale dependence of the Yukawa interaction.
\end{itemize}

The paper is organized as follows: In Secs.\ \ref{sec:models} and \ref{sec:methods}, we introduce the basic 
models, methods, and concepts that are used in this work. To this end, 
Secs.\ \ref{sec:LSM} and \ref{sec:NLSM} briefly review the $O(4)$ LSM, the NLSM, as 
well as the nonlinear formalism. Afterwards, Sec.\ \ref{sec:FRG} focuses on the 
higher-derivative interactions in terms of the FRG truncation. In Sec.\ \ref{sec:effpion}, 
we transform the linearly realized effective action of the QMM into its nonlinear counterpart. 
Finally, Sec.\ \ref{sec:results} presents the numerical results of the FRG flow within the 
linear QMM, cf.\ Sec.\ \ref{sec:higher_order}, and the computed low-energy couplings of the 
associated nonlinear model, cf.\ Sec.\ \ref{sec:LECs}. The conclusions of this work as well 
as an outlook for further investigations is given in Sec.\ \ref{sec:summary}.

\section{Models}
\label{sec:models}

In this section, we briefly summarize the most important features of the linear QMM
and the nonlinear model with pionic degrees of freedom.

\subsection{The linear Quark-Meson Model}
\label{sec:LSM}

As already mentioned in the introduction, the simplest model based on a linear 
realization of chiral symmetry is given by the $O(4)$ LSM. The basic object of this 
model is the four-dimensional Euclidean field-space vector
\begin{equation}
	\varphi = \begin{pmatrix} \vec{\pi} \\ \sigma \end{pmatrix}. \label{eq:phi}
\end{equation}
This vector constitutes the fundamental representation of $O(4)$ and, hence, transforms as
\begin{equation}
	\varphi \xrightarrow{O(4)} O\varphi, \quad O \in O(4).
	\label{eq:phitrafo}
\end{equation}  
The Lagrangian of the $O(4)$ LSM is then constructed as
\begin{equation}
	\mathcal{L}_{\text{LSM}} = \frac{1}{2}\left(\partial_{\mu}\varphi\right)\cdot\partial^{\mu}\varphi 	
	- \frac{m_{0}^{2}}{2}\varphi\cdot\varphi - \frac{\lambda}{4}\left(\varphi\cdot\varphi\right)^{2}.
	\label{eq:lagrangianLSM}
\end{equation} 

The QMM is obtained from the above Lagrangian by including quarks in a chirally invariant way,
\begin{IEEEeqnarray}{rCl}
	\mathcal{L}_{\text{QMM}} & = & \frac{1}{2}\left(\partial_{\mu}\varphi\right)\cdot\partial^{\mu}\varphi 
	- \frac{m_{0}^{2}}{2}\varphi\cdot\varphi - \frac{\lambda}{4}\left(\varphi\cdot\varphi\right)^{2} + 
	h_{\text{ESB}}\sigma \nonumber\\
	& & +\, \bar{\psi}\left(i\gamma^{\mu}\partial_{\mu} - y\Phi_{5}\right)\psi, \label{eq:lagrangianQMM}
\end{IEEEeqnarray}
with 
\begin{equation}
	\Phi_{5} = \sigma t_{0} + i\gamma_{5}\vec{\pi}\cdot\vec{t},
\end{equation}
where $t_{0} = \mathbbmss{1}_{2}/2$ and $\vec{t} = \vec{\tau}/2$. Here, $\vec{\tau}$ 
denotes the usual vector of the Pauli matrices. The normalization of the generators 
is chosen such that $\mathrm{tr}\left(t_{a}t_{b}\right) = \delta_{ab}/2$, $a,b = 0,\ldots,3$. 
The additional term $\sim \!h_{\text{ESB}}\sigma$ in the above Lagrangian describes the explicit 
breaking of chiral symmetry (ESB) by tilting the mesonic potential into the direction 
of the $\sigma$ field. 

In addition to the ESB, also the spontaneous breaking of chiral symmetry
has to be modelled. The latter is signaled by a nonvanishing order parameter identified 
with the vacuum expectation value $\sigma_{0}$ of the $\sigma$ field. This order 
parameter is typically introduced by shifting the $\sigma$ field according to
\begin{equation}
	\sigma \rightarrow \sigma_{0} + \sigma.
\end{equation}

\subsection{The nonlinear model}
\label{sec:NLSM}

In contrast to the LSM, described by Eq.\ (\ref{eq:lagrangianLSM}), the field space 
of the associated nonlinear model is not given by the four-dimensional Euclidean space, 
but by a three-dimensional submanifold \cite{Meetz:1969as}. This field space arises as a consequence 
of the pattern of spontaneous symmetry breaking. It is defined by the degenerate vacua and is 
usually denoted as the vacuum manifold.
Since the coordinates of this space are in one-to-one correspondence with the pNGBs, the basic 
degrees of freedom of the NLSM are given by the three pion fields.

For the following discussion, we only consider the $SO(4)$ subgroup of $O(4)$, such that the 
vacuum manifold of the LSM is given by the space of (left) cosets $SO(4)/SO(3)$, cf.\ Refs.\ 
\cite{Weinberg:1968de, Coleman:1969sm, Callan:1969sn}. Using a representative of this coset 
space, in the following denoted as $\Sigma\left(\zeta\right)$, it is possible to construct 
the so-called Maurer-Cartan form $\alpha_{\mu}(\zeta)$ as
\begin{equation}
	\alpha_{\mu}\left(\zeta\right) = \Sigma^{-1}\left(\zeta\right)\partial_{\mu}\Sigma
	\left(\zeta\right).
	\label{eq:MCform1}
\end{equation}
It should be emphasized that the coordinates of the coset space, $\zeta^{\alpha}$, 
$\alpha = 1,2,3$, are usually not exactly identical to the pion fields, but directly related 
to them. The Maurer-Cartan form is defined in the Lie algebra $\mathfrak{so}(4)$ and can 
therefore be expanded as
\begin{equation}
	\alpha_{\mu}\left(\zeta\right) = ie^{a}_{\mu}\left(\zeta\right)x_{a} + i\omega^{i}_{\mu}
	\left(\zeta\right)s_{i},
	\label{eq:MCform2}
\end{equation}
with coefficients
\begin{equation}
	e^{a}_{\mu}\left(\zeta\right) = e^{\;\;a}_{\alpha}\left(\zeta\right)\partial_{\mu}\zeta^{\alpha}, 
	\qquad \omega^{i}_{\mu}\left(\zeta\right) = \omega^{\;\;i}_{\alpha}\left(\zeta\right)\partial_{\mu}
	\zeta^{\alpha},
	\label{eq:MCform3}
\end{equation}
where $x_{a}$, $a = 1, 2, 3$, denotes the coset generators and $s_{i}$, $i = 1, 2, 3$, 
those of the unbroken $SO(3)$ subgroup. The coefficients $e^{\;\;a}_{\alpha}(\zeta)$
define a frame on $SO(4)/SO(3)$ and the related metric reads
\begin{equation}
	g_{\alpha\beta}\left(\zeta\right) = \delta_{ab}\, e^{\;\;a}_{\alpha}\left(\zeta\right)
	e^{\;\;b}_{\beta}\left(\zeta\right),
	\label{eq:metric}
\end{equation}
where $\alpha, \beta = 1,2,3$ represent curved coset indices.

The simplest Lagrangian that can be constructed from the above objects is given by
\begin{equation}
	\mathcal{L}_{\text{NLSM}} = \frac{1}{2}F_{ab}\, e^{a}_{\mu}\left(\zeta\right)e^{b,\mu}		
	\left(\zeta\right),
	\label{eq:NLSMlag}
\end{equation} 
where the real-valued matrix $F$ is needed for dimensional reasons. This Lagrangian
is usually called the $SO(4)$ NLSM and contains the pion self-interaction 
terms to arbitrary order in the fields with at most two spacetime derivatives. 

In Appendix \ref{sec:trafoprops}, we review the general transformation properties of the coset representative
and the Maurer-Cartan form. In addition, we present the transformation behavior of the 
nonlinear pion fields for an explicit choice of coordinates.

\section{Methods}
\label{sec:methods}

In this section, we discuss the calculation of the linearly realized effective action of the QMM 
and its relation to the nonlinear model featuring only pionic degrees of freedom.

A necessary prerequisite for the determination of the low-energy couplings of 
the nonlinear model from the effective action in terms of linearly realized pion fields 
is the integration of all nonpionic QCD fluctuations, and, in particular, the quark loops. 
Moreover, such a determination of the low-energy couplings is only meaningful at scales, 
where fluctuations of nonpionic fields have already decoupled from the dynamics.

Before going on, we want to point out that the quark fluctuations of the QMM simulate full QCD dynamics. 
In contrast to the purely mesonic linear model, including these quark fluctuations entails the qualitatively 
correct decoupling of the mesonic degrees of freedom above the scale of chiral symmetry breaking. 
Quantitatively, pure quark fluctuations are not capable of fully capturing QCD dynamics and, e.g., 
this decoupling happens too slowly \cite{Braun:2014ata, 
Mitter:2014wpa, Cyrol:2017ewj}. An investigation of the effect of full QCD dynamics on the determination 
of the low-energy couplings is deferred to future work.

\subsection{Effective action from the\\Functional Renormalization Group}
\label{sec:FRG}

The FRG is a nonperturbative continuum method that formulates the integration of quantum 
fluctuations in terms of the RG-scale ($k$) dependence of the effective average 
action $\Gamma_{k}$, which smoothly interpolates between the renormalized
classical action $S$ at the ultraviolet (UV) cutoff scale $\Lambda$, $\Gamma_{k \rightarrow \Lambda} 
= S$, and the quantum effective action $\Gamma$ in the infrared (IR), $\Gamma_{k \rightarrow 0}
= \Gamma$. The action $\Gamma$ is the generating functional for all one-particle 
irreducible vertex functions.

The scale evolution of $\Gamma_{k}$ from the UV to the IR is described by the 
Wetterich equation \cite{Wetterich:1992yh},
\begin{IEEEeqnarray}{rCl}
	\partial_{k}\Gamma_{k} & = & \frac{1}{2}\tr\left[
	\partial_{k} R_{k}\left(
	\Gamma^{(2)}_{k} + R_{k}\right)^{-1}\right] \nonumber\\
	& = & \frac{1}{2} \! \!
	\vcenter{\hbox{
	\begin{pspicture}(1.5,2.0)
		\psarc[linewidth=0.02](0.75,1.0){0.6}{113}{67}
		\pscircle[linewidth=0.03](0.75,1.6){0.25}
		\psline[linewidth=0.03](0.75,1.6)(0.92,1.77)
		\psline[linewidth=0.03](0.75,1.6)(0.58,1.77)
		\psline[linewidth=0.03](0.75,1.6)(0.58,1.43)
		\psline[linewidth=0.03](0.75,1.6)(0.92,1.43)
	\end{pspicture}
	}} \! \! , \label{eq:Wetterich} 
\end{IEEEeqnarray}
where the second line introduces the common diagrammatical notation
\begin{equation}
	\left(\Gamma^{(2)}_{k} + R_{k}\right)^{-1} = 
	\! \! \! \vcenter{\hbox{
	\begin{pspicture}[showgrid=false](1.5,0.5)
		\psline[linewidth=0.02](0.1,0.25)(1.4,0.25)
	\end{pspicture}
	}} \! \! ,\quad
	\partial_{k}R_{k} = \! \! \!
	\vcenter{\hbox{
	\begin{pspicture}[showgrid=false](1.5,0.8)
		\psline[linewidth=0.02](0.1,0.4)(0.5,0.4)
		\psline[linewidth=0.02](1.0,0.4)(1.4,0.4)
		\pscircle[linewidth=0.03](0.75,0.4){0.25}
		\psline[linewidth=0.03](0.75,0.4)(0.92,0.57)
		\psline[linewidth=0.03](0.75,0.4)(0.58,0.57)
		\psline[linewidth=0.03](0.75,0.4)(0.58,0.23)
		\psline[linewidth=0.03](0.75,0.4)(0.92,0.23)
	\end{pspicture}
	}} \! \! . \label{eq:rep}
\end{equation}
The above flow equation contains the regulator function $R_{k}$, which gives 
an additional mass contribution for low-energy modes. This means that it 
effectively acts as an IR cutoff separating those soft modes from the integration 
process. By successively lowering the scale $k$, the effective average action
$\Gamma_{k}$ includes increasingly more fluctuations and, in the limit $k \rightarrow 0$, 
all quantum fluctuations are integrated out.

In order to compute $\Gamma_{k}$, one has to truncate the system of
vertex functions, since the right-hand side of Eq.\ (\ref{eq:Wetterich}) involves
the two-point function $\Gamma_{k}^{(2)}$, which, itself, couples through the flow 
to higher $n$-point functions. A typical truncation scheme for these vertex functions 
is given by a derivative expansion, which we will focus on in the following; 
cf.\ also the introductory Refs.\ \cite{Bonini:1992vh, Ellwanger:1993mw, Morris:1993qb, 
Bagnuls:2000ae, Berges:2000ew, Pawlowski:2005xe, Blaizot:2006rj, Gies:2006wv, Schaefer:2006sr, 
Kopietz:2010zz, vonSmekal:2012vx}.

Along the lines of our previous study \cite{Eser:2018jqo}, we choose the following 
(Euclidean) truncation for the linearly realized $O(4)$ QMM based on Eq.\ (\ref{eq:lagrangianQMM}) 
in Sec.\ \ref{sec:LSM} and introduce the $k$-dependent higher-derivative couplings $C_{2,k}$ 
and $Z_{2,k}$, as well as $C_{i,k}$, $i = 3,\ldots , 8$:
\begin{IEEEeqnarray}{rCl}
	\Gamma_{k} & = & \int_{x}\bigg\lbrace
	\frac{Z_{k}}{2}
	\left(\partial_{\mu}\varphi\right) \cdot 
	\partial_{\mu}\varphi + U_{k} - h_{\mathrm{ESB}}\sigma 
	\nonumber\\
	& & \qquad +\, C_{2,k}\left(\varphi \cdot 
	\partial_{\mu}\varphi\right)^{2}
	+ Z_{2,k}\, \varphi^{2}\left(\partial_{\mu}\varphi\right)
	\cdot \partial_{\mu}\varphi
	\nonumber\\
	& & \qquad -\, C_{3,k} 
	\left[\left(\partial_{\mu}\varphi\right)
	\cdot \partial_{\mu}\varphi\right]^{2}
	- C_{4,k} \left[\left(\partial_{\mu}\varphi\right)
	\cdot \partial_{\nu}\varphi\right]^{2} \nonumber\\
	& & \qquad -\, C_{5,k} \,
	\varphi \cdot \left(\partial_{\mu}\partial_{\mu}\varphi\right)
	\left(\partial_{\nu}\varphi\right) \cdot \partial_{\nu}\varphi
	\nonumber\\
	& & \qquad -\, C_{6,k} \,
	\varphi^{2} \left(\partial_{\mu}\partial_{\nu}\varphi\right)
	\cdot \partial_{\mu}\partial_{\nu}\varphi
	\nonumber\\
	& & \qquad -\, C_{7,k} 
	\left(\varphi \cdot \partial_{\mu}\partial_{\mu}\varphi\right)^{2}
	- C_{8,k} \,
	\varphi^{2}\left(\partial_{\mu}\partial_{\mu}\varphi\right)^{2}
	\nonumber\\
	& & \qquad +\, \bar{\psi}\left(Z_{k}^{\psi}\gamma_{\mu}
	\partial_{\mu}+ 
	y_{k} \Phi_{5}\right)
	\psi\bigg\rbrace .\label{eq:truncation}
\end{IEEEeqnarray}
This truncation constitutes a derivative expansion up to order 
$\mathcal{O}(\partial^{4})$. It is well beyond the LPA and its minimal extension
known as the LPA'. The first would only consider a scale-dependent effective
potential $U_{k}$, which is a function of the $O(4)$ invariant $\rho$, 
\begin{equation}
	\rho \equiv \varphi \cdot \varphi = \vec{\pi}^{2} + \sigma^{2},
\end{equation}
while the latter would also take into account the scaling of 
the field variables by means of the flow of the wave-function renormalization 
factors for bosons and fermions, $Z_{k}$ and $Z_{k}^{\psi}$. In both of these two approximations,
the higher-derivative couplings would be absent, i.e., they are set to zero.
In the light of Ref.\ \cite{Eser:2018jqo}, the momentum-independent coupling 
$C_{1,k}(\varphi \cdot \varphi)^{2}$ is omitted in the above equation. 
It is to be indentified with the quartic interaction term of the effective potential.

The parameter $h_{\mathrm{ESB}} \neq 0$ explicitly breaks the $O(4)$ symmetry, 
as mentioned in the last section, and remains $k$ independent. For the scale-dependent 
factors $Z_{k}$, $Z_{k}^{\psi}$, and the Yukawa interaction $y_{k}$ we suppress 
a general field dependence. The couplings $\lbrace C_{2,k}, Z_{2,k}\rbrace$ as well as 
$\lbrace C_{i,k} \rbrace$, $i = 3,\ldots , 8$, beyond the LPA' form a complete set of terms 
of order $\mathcal{O}(\varphi^{4}, \partial^{2})$ and $\mathcal{O}(\varphi^{4}, 
\partial^{4})$, respectively. As an extension of Ref.\ \cite{Eser:2018jqo} 
[cf.\ Eq.\ (20) therein], the low-energy couplings in Eq.\ (\ref{eq:truncation}) 
now include momentum-de\-pen\-dent $\sigma\pi$ and $\sigma$ self-interactions.

On the level of the two-point functions $\Gamma_{k}^{(2)}$, we define different 
effective wave-function renormalization factors for the $\sigma$
and $\pi$ fields,
\begin{IEEEeqnarray}{rCl}
	Z_{k}^{\sigma} & = & Z_{k} + 2 \sigma^{2} \left(Z_{2,k} + C_{2,k}\right) \nonumber\\ 
	& & - 2 \sigma^{2} p^{2} \left(C_{6,k} + C_{7,k} + C_{8,k}\right), \label{eq:Zsigma}\\
	Z_{k}^{\pi} & = & Z_{k} + 2 \sigma^{2} Z_{2,k} - 2 \sigma^{2} p^{2} 
	\left(C_{6,k} + C_{8,k}\right), \label{eq:Zpi}
\end{IEEEeqnarray}
where $p$ is the external momentum from the functional derivatives with respect 
to the fields. The corrections to $Z_{k}$ in these definitions 
obviously arise from the presence of the higher-derivative couplings. 
It should be noted that the distinction between $Z_{k}^{\sigma}$ and $Z_{k}^{\pi}$ is not 
in contradiction to the $O(4)$ symmetry of the model. As soon as the $\sigma$ field acquires
a nonvanishing expectation value, $\sigma \neq 0$, the wave-function renormalizations will 
naturally split.

For later reference, we define the following renormalized quantities based on 
the wave-function renormalization factors $Z_{k}^{\pi}$ and $Z_{k}^{\psi}$: 
\begin{IEEEeqnarray}{rCl}
	\tilde{\sigma} & = & \sqrt{Z_{k}^{\pi}} \sigma, \quad
	\tilde{\vec{\pi}} = \sqrt{Z_{k}^{\pi}} \vec{\pi}, \nonumber\\
	\tilde{\psi} & = & \sqrt{Z_{k}^{\psi}} \psi, \quad
	\tilde{\bar{\psi}} = \sqrt{Z_{k}^{\psi}} \bar{\psi}, \nonumber\\
	\tilde{h}_{\mathrm{ESB}} & = & \frac{h_{\mathrm{ESB}}}{\sqrt{Z_{k}^{\pi}}}, \quad
	\tilde{y}_{k} = \frac{y_{k}}{Z_{k}^{\psi} \sqrt{Z_{k}^{\pi}}}, \nonumber\\
	\tilde{C}_{i,k} & = & \frac{C_{i,k}}{\left(Z_{k}^{\pi}\right)^{2}}, \quad
	i = 1, \ldots , 8, \quad
	\tilde{Z}_{2,k} = \frac{Z_{2,k}}{\left(Z_{k}^{\pi}\right)^{2}} ,\label{eq:rquantities}
\end{IEEEeqnarray}
where we will, for simplicity, evaluate the above wave-function renormalization factors 
at vanishing external momentum, $p=0$; cf.\ the discussion in Appendix \ref{sec:floweqns}.
By choosing $Z_{k}^{\pi}$ for both bosonic fields in the definitions (\ref{eq:rquantities})
one directly obtains the correct renormalization factors in the nonlinear model, 
as presented in the next section.

The flow equations for all scale-dependent quantities in truncation (\ref{eq:truncation})
are obtained by projecting the functional derivatives of Eq.\ (\ref{eq:Wetterich}) (the flows
of vertex functions) onto the respective coupling. The resulting expressions for these equations 
and further technical aspects, such as the regulator functions, are shown in Appendix \ref{sec:floweqns}.

Finally, the integration of the coupled system of differential equations then allows for a computation 
of the quantities in the linearly realized effective average action (\ref{eq:truncation}) in the IR limit, 
$\Gamma_{k \rightarrow 0} = \Gamma$; see Appendix \ref{sec:solv_floweqns} for details.

\subsection{Effective pion action}
\label{sec:effpion}

A physically meaningful transition from the effective action to the nonlinear model as effective
low-energy theory requires that all fluctuations, except for those of the pions, should have 
decoupled from the dynamics at the energy-momentum scale, where this transition is to be realized. 
The low-energy limit of the effective action (\ref{eq:truncation}), expressed in terms of the
renormalized quantities (\ref{eq:rquantities}), is then constructed by integrating out all (already 
decoupled) fields,
\begin{equation}
	\Gamma_{k}\left[\tilde{\sigma},\tilde{\vec{\pi}},\tilde{\bar{\psi}},\tilde{\psi}\right] \quad		
	\longrightarrow \quad \Gamma_{k}\left[\tilde{\Pi}\right].
\end{equation}
The vector $\tilde{\Pi}=(\tilde{\Pi}^{1},\tilde{\Pi}^{2},\tilde{\Pi}^{3})$ represents the 
renormalized nonlinear pNGB fields, which will be defined below, and the symbol $\Gamma_k$ is kept 
for the resulting action.

The quark fields are immediately dropped from the effective action, similarly 
to the investigations in Refs.\ \cite{Divotgey:2016pst, Eser:2018jqo}, since this integration
process is restricted to (at most) tree-level diagrams. As a consequence, $\Gamma_{k}$ reduces to the 
effective action of the usual LSM, modified by the higher-derivative couplings.

As already mentioned in Sec.\ \ref{sec:NLSM}, the $SO(4)$ NLSM is defined on the 
coset space $SO(4)/SO(3)$, which is diffeomorphic to the three-sphere $S^{3}$. 
The explicit diffeomorphism is given by
\begin{equation}
	\tilde{\varphi} \equiv \sqrt{Z_{k}^{\pi}}\varphi = \Sigma\bigl(\tilde{\zeta}\bigr)\tilde{\phi},
	\label{eq:iso}
\end{equation}
with $\tilde{\phi} = (\vec{0}, \tilde{\theta})$, where $\tilde{\theta} 
= \sqrt{\tilde{\varphi}^{2}}$ defines the radius of the three-sphere. For the 
time being, we keep the field $\tilde{\theta}$ and allow fluctuations in radial 
direction. 

The coordinates $\tilde{\zeta}^{a}$ parametrizing the coset representative 
$\Sigma(\tilde{\zeta})$ are chosen as stereographic projections,
\begin{equation}
    \tilde{\zeta}^{a} = \frac{\tilde{\pi}^{a}}{\tilde{\theta} + \tilde{\sigma}}, 
    \quad a=1,2,3,
    \label{eq:coords}
\end{equation}
where $\tilde{\pi}^{a}$ and $\tilde{\sigma}$ are the renormalized Euclidean 
coordinates of the linear QMM, cf.\ Sec.\ \ref{sec:FRG}. An explicit 
parametrization of the coset representative then reads
\begin{equation}
	\Sigma\bigl(\tilde{\zeta}\bigr) = \begin{pmatrix} \delta^{a}_{\;\; b} - 
	\frac{2\tilde{\zeta}^{a}		
	\tilde{\zeta}_{b}}{1 + \tilde{\zeta}^{2}} && \frac{2\tilde{\zeta}^{a}}{1 + \tilde{\zeta}
	^{2}} \\[0.2cm]
    -\frac{2\tilde{\zeta}_{b}}{1 + \tilde{\zeta}^{2}} && \frac{1 - \tilde{\zeta}^{2}}{1 + 
    \tilde{\zeta}^{2}}
	\end{pmatrix}. \label{eq:cosetrep}
\end{equation}
The coefficients of the Maurer-Cartan form proportional to the broken generators as well
as the metric in Eq.\ (\ref{eq:metric}) are thus evaluated as
\begin{equation}
	e^{a}_{\mu}\bigl(\tilde{\zeta}\bigr) \equiv e^{\;\;a}_{\alpha}\bigl(\tilde{\zeta}\bigr)
	\partial_{\mu}\tilde{\zeta}^{\alpha} = 
	\frac{2\delta^{\;\;a}_{\alpha}}{1 + \tilde{\zeta}^{2}}\partial_{\mu}\tilde{\zeta}^{\alpha},
	\label{eq:mcframe}
\end{equation}
and
\begin{equation}
	g_{\alpha\beta} \equiv g_{\alpha\beta}\bigl(\tilde{\zeta}\bigr) = \frac{4\delta_{\alpha\beta}}{\bigl(1 + 
	\tilde{\zeta}^{2}\bigr)^{2}},
\end{equation}
respectively.
 
The transition of the LSM to its associated NLSM is now realized by inserting 
Eq.\ (\ref{eq:iso}) into Eq.\ (\ref{eq:truncation}) and identifying the $\tilde{\theta}$
field with the pion decay constant $f_{\pi}$ [PCAC relation; cf.\ Ref.\ \cite{Eser:2018jqo}],
\begin{equation}
	\tilde{\theta} = f_{\pi}.
\end{equation}
This step fixes the radial fluctuations $\tilde{\theta}$ to a constant radius
and eliminates the $\tilde{\theta}$ field from the effective action. The resulting nonlinear 
model can then be written as
\begin{IEEEeqnarray}{rCl}
	\Gamma_{k} & = & \int_{x} \bigg\lbrace \frac{f^{2}_{\pi}}{2}g_{\alpha\beta}\bigl(\nabla_{\mu}
	\tilde{\zeta}^{\alpha}\bigr)\nabla_{\mu}\tilde{\zeta}^{\beta} \nonumber\\
	& & \quad\quad -\, \bigl(\tilde{C}_{6,k}+\tilde{C}_{8,k}\bigr) \, f_{\pi}^{4} \, 
	 g_{\alpha\beta}\bigl(\nabla_{\mu}\nabla_{\mu}
	\tilde{\zeta}^{\alpha}\bigr)
	\bigl(\nabla_{\nu}\nabla_{\nu}\tilde{\zeta}^{\beta}\bigr) \nonumber\\
    & & \quad\quad -\, \bigl(\tilde{C}_{3,k} - \tilde{C}_{5,k} + \tilde{C}_{6,k} + \tilde{C}_{7,k} 
    + \tilde{C}_{8,k}\bigr)\, f^{4}_{\pi} \nonumber\\
    & & \qquad\qquad \times \, g_{\alpha\beta}g_{\gamma\delta}\bigl(\nabla_{\mu}
    \tilde{\zeta}^{\alpha}\bigr)\bigl(\nabla_{\mu}\tilde{\zeta}^{\beta}\bigr)\bigl(\nabla_{\nu}
    \tilde{\zeta}^{\gamma}\bigr)\nabla_{\nu}\tilde{\zeta}^{\delta} \nonumber\\ 
	& & \quad\quad -\, \tilde{C}_{4,k} \, f^{4}_{\pi} \,
	g_{\alpha\beta}g_{\gamma\delta}\bigl(\nabla_{\mu}
    \tilde{\zeta}^{\alpha}\bigr)\bigl(\nabla_{\nu}\tilde{\zeta}^{\beta}\bigr)\bigl(\nabla_{\mu}
    \tilde{\zeta}^{\gamma}\bigr)\nabla_{\nu}\tilde{\zeta}^{\delta} \nonumber\\ 
	& & \quad\quad -\, \tilde{h}_{\text{ESB}}\, f_{\pi}\frac{1 - \tilde{\zeta}^{2}}{1 + \tilde{\zeta}^{2}}
	\bigg\rbrace, \label{eq:integration}
\end{IEEEeqnarray}
with $\nabla_{\mu}\tilde{\zeta}^{\alpha} \equiv \partial_{\mu}\tilde{\zeta}^{\alpha}$.
The action of the covariant derivative $\nabla_{\mu}$ on a vector $\partial_{\nu}\tilde{\zeta}^{\alpha}$ is defined as
\begin{equation}
	\nabla_{\mu}\nabla_{\nu}\tilde{\zeta}^{\alpha} \equiv	
	\nabla_{\mu}\partial_{\nu}\tilde{\zeta}^{\alpha} = 
	\partial_{\mu}\partial_{\nu}\tilde{\zeta}^{\alpha} +
	\Gamma^{\alpha}_{\;\;\beta\gamma}\bigl(\partial_{\mu}\tilde{\zeta}^{\beta}\bigr)\partial_{\nu}
	\tilde{\zeta}^{\gamma}.
\end{equation}
In stereographic coordinates, the above Christoffel symbols $\Gamma^{\alpha}_{\;\;\beta\gamma}$ read
\begin{equation}
	\Gamma^{\alpha}_{\;\;\beta\gamma} = \frac{2}{1 + \tilde{\zeta}^{2}}\left(
	-\delta^{\alpha}_{\;\;\gamma}\tilde{\zeta}_{\beta} - \delta^{\alpha}_{\;\;\beta}\tilde{\zeta}_{\gamma}
	+ \delta_{\beta\gamma}\tilde{\zeta}^{\alpha}\right).
\end{equation}
The effective action (\ref{eq:integration}) resembles Eq.\ (\ref{eq:NLSMlag}) 
with $F_{ab} = f_{\pi}^{2}\delta_{ab}$ for these specific coordinates. 
Furthermore, it features corrections from the ESB term and the higher-derivative 
couplings. Its general form is consistent with the studies on nonlinear 
sigma models in Refs.\ \cite{Percacci:2009fh, Flore:2012ma}.

In order to obtain a canonically normalized kinetic term for the pNGB fields, we introduce a 
field redefinition according to
\begin{equation}
	\tilde{\zeta}^{a} \rightarrow \frac{\tilde{\Pi}^{a}}{2f_{\pi}}, \quad a = 1, 2, 3.
\end{equation}
Expanding all quantities up to fourth order in the new field variables $\tilde{\Pi}^{a}$, 
the above effective action becomes
\begin{IEEEeqnarray}{rCl}
	\Gamma_{k} & = & \int_{x} \biggl\lbrace \frac{1}{2}
	\left(\partial_{\mu}\tilde{\Pi}_{a}\right)\partial_{\mu}\tilde{\Pi}^{a} + \frac{1}{2}
	\tilde{\mathcal{M}}^{2}_{\Pi,k}\, \tilde{\Pi}_{a}\tilde{\Pi}^{a} \nonumber\\
	& & \quad\quad -\, \tilde{\mathcal{C}}_{1,k}\left(\tilde{\Pi}_{a}\tilde{\Pi}^{a}
	\right)^{2} + \tilde{\mathcal{Z}}_{2,k}\, \tilde{\Pi}_{a}\tilde{\Pi}^{a}
	\left(\partial_{\mu}\tilde{\Pi}_{b}\right)\partial_{\mu}\tilde{\Pi}^{b} \nonumber\\
	& & \quad\quad -\, \tilde{\mathcal{C}}_{3,k}\left[\left(\partial_{\mu}\tilde{\Pi}_{a}\right)
	\partial_{\mu}\tilde{\Pi}^{a}\right]^{2} \nonumber\\
	& & \quad\quad -\, \tilde{\mathcal{C}}_{4,k}\left[\left(\partial_{\mu}\tilde{\Pi}_{a}
	\right)\partial_{\nu}\tilde{\Pi}^{a}\right]^{2} \nonumber\\
	& & \quad\quad -\, \tilde{\mathcal{C}}_{5,k}\, \tilde{\Pi}_{a}\left(\partial_{\mu}\partial_{\mu}
	\tilde{\Pi}^{a}\right)\left(\partial_{\nu}\tilde{\Pi}_{b}\right)\partial_{\nu}
	\tilde{\Pi}^{b} \nonumber\\
	& & \quad\quad -\, \tilde{\mathcal{C}}_{6,k}\, \tilde{\Pi}_{a}\tilde{\Pi}^{a}
	\left(\partial_{\mu}\partial_{\nu}\tilde{\Pi}_{b}\right)\partial_{\mu}\partial_{\nu}\tilde{\Pi}^{b} 
	\nonumber \\
	& & \quad\quad -\, \tilde{\mathcal{C}}_{8,k}\, \tilde{\Pi}_{a}\tilde{\Pi}^{a}
	\left(\partial_{\mu}\partial_{\mu}\tilde{\Pi}_{b}\right)\partial_{\nu}\partial_{\nu}\tilde{\Pi}^{b}
	\biggr\rbrace ,
	\label{eq:finalaction}
\end{IEEEeqnarray}
where we defined the squared mass of the $\tilde{\Pi}^{a}$ fields,
\begin{equation}
	\tilde{\mathcal{M}}^{2}_{\Pi,k} = \frac{\tilde{h}_{\text{ESB}}}{f_{\pi}},
\end{equation}
as well as the low-energy couplings
\begin{IEEEeqnarray}{rCl}
	\tilde{\mathcal{C}}_{1,k} & = & \frac{\tilde{\mathcal{M}}^{2}_{\Pi,k}}{8f^{2}_{\pi}}, \nonumber\\
	\tilde{\mathcal{Z}}_{2,k} & = &-\frac{1}{4f^{2}_{\pi}}, \nonumber\\
	\tilde{\mathcal{C}}_{3,k} & = & \tilde{C}_{3,k}-\tilde{C}_{5,k}+\tilde{C}_{7,k}
	+ 2\bigl(\tilde{C}_{6,k}+\tilde{C}_{8,k}\bigr), \nonumber \\
	\tilde{\mathcal{C}}_{4,k} & = & \tilde{C}_{4,k}, \nonumber \\
	\tilde{\mathcal{C}}_{5,k} & = & 2\bigl(\tilde{C}_{6,k}+\tilde{C}_{8,k}\bigr), \nonumber \\
	\tilde{\mathcal{C}}_{6,k} & = & -\, \tilde{C}_{6,k}-\tilde{C}_{8,k}, \nonumber \\
	\tilde{\mathcal{C}}_{8,k} & = & \frac{1}{2}\bigl(\tilde{C}_{6,k} + \tilde{C}_{8,k}\bigr).
	\label{eq:LECrelations}
\end{IEEEeqnarray} 
It should be underlined that Eq.\ (\ref{eq:finalaction}) is calculated from 
Eq.\ (\ref{eq:integration}) by integrating several terms by parts in order to 
reobtain the term structures of the derivative expansion (\ref{eq:truncation}).
\begin{figure*}[t!]
	\centering
		\includegraphics{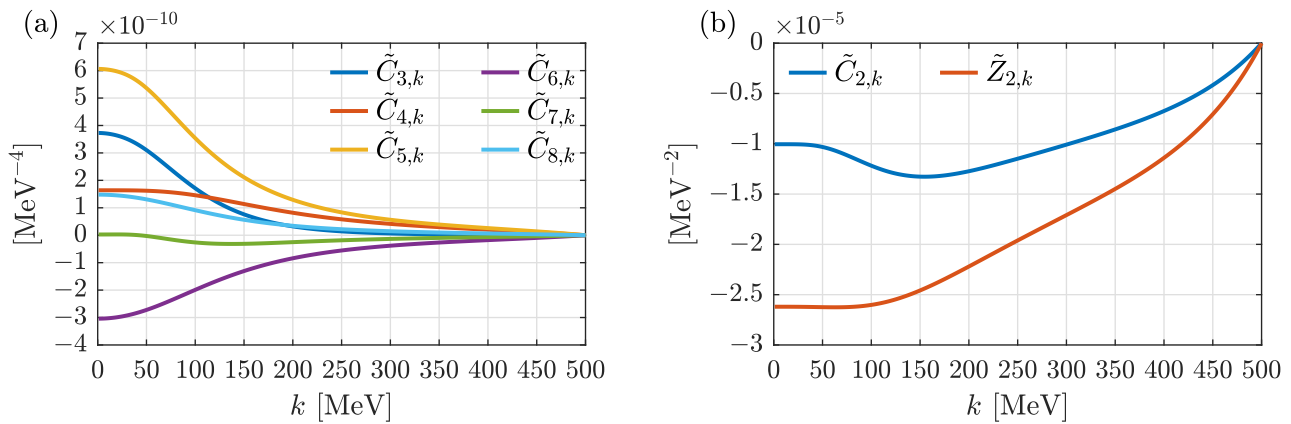}
	\caption{Higher-derivative couplings of the linear QMM. (a)
	Scale evolution of the renormalized couplings $\tilde{C}_{i,k}$, $i = 3, \ldots , 8$. 
	(b) Scale evolution of the renormalized couplings $\tilde{C}_{2,k}$ and $\tilde{Z}_{2,k}$;
    $k_{\mathrm{IR}} = 1\ \mathrm{MeV}$.}
	\label{fig:c}
\end{figure*}

Equation (\ref{eq:finalaction}) is one of the central results of this paper. It is the nonlinear
counterpart of the linear QMM and, as repeatedly pointed out, has to be understood as its 
low-energy limit. Moreover, we obtained relations between the low-energy couplings in the linear
and the nonlinear model, cf.\ Eq.\ (\ref{eq:LECrelations}).

It is remarkable that the geometrical constraint of fixing the $\tilde{\theta}$ field
to the constant radius of the three-sphere restricts the number of possible couplings of
order $\mathcal{O}(\partial^{2})$ as well as $\mathcal{O}(\partial^{4})$. In the chosen set of 
coordinates, only the analogue of $\tilde{Z}_{2,k}$, the coupling $\tilde{\mathcal{Z}}_{2,k}$ out of the terms 
of order $\mathcal{O}(\partial^{2})$, ``survives'' in the nonlinear framework, while the interaction 
$\sim \tilde{C}_{2,k}$ vanishes. The same holds true for the couplings $\tilde{C}_{3,k}$, $\tilde{C}_{4,k}$, 
$\tilde{C}_{5,k}$, $\tilde{C}_{6,k}$, and $\tilde{C}_{8,k}$ in the case of the terms of order 
$\mathcal{O}(\partial^{4})$. Besides, it obviously arises a linear dependence in the nonlinear model
between the couplings $\tilde{\mathcal{C}}_{5,k}$, $\tilde{\mathcal{C}}_{6,k}$, and 
$\tilde{\mathcal{C}}_{8,k}$ after fixing the $\tilde{\theta}$ field, see once more 
Eq.\ (\ref{eq:LECrelations}).

We observe that the mo\-men\-tum-in\-de\-pen\-dent quar\-tic coupling 
$\tilde{C}_{1,k}$ and the effective potential $U_{k}$, in general, do 
not enter the low-energy limit. Also, the couplings $\tilde{C}_{2,k}$ and $\tilde{Z}_{2,k}$ are 
irrelevant for the results in the nonlinear effective pion action.
They only indirectly influence the result through the integration process of the system
of flow equations. Furthermore, the couplings $\tilde{\mathcal{C}}_{1,k}$ and $\tilde{\mathcal{Z}}_{2,k}$ are 
identical to the analytical results for the respective terms in the ChPT Lagrangian formulated 
in stereographic coordinates, as can be easily deduced from Ref.\ \cite{Divotgey:2016pst}. In fact,
within the geometrical constraints on the vacuum manifold, the coupling $\tilde{\mathcal{Z}}_{2,k}$ 
is only a function of the pion decay constant $f_{\pi}$.

\section{Numerical Results and Discussion}
\label{sec:results}

We now present the numerical results for the higher-derivative interactions 
as obtained from the linear QMM, renormalized at a hadronic cutoff scale of
$\Lambda = 500\ \mathrm{MeV}$, as well as the derived low-energy couplings of the
nonlinear model. All additional constituents of truncation (\ref{eq:truncation}) 
and details on their calculation within the FRG framework are described in 
Appendix \ref{sec:solv_floweqns}.

\subsection{Linear model: Higher-derivative couplings}
\label{sec:higher_order}

The results for the higher-derivative pion couplings of the linear QMM are shown in 
Fig.\ \ref{fig:c}. The subfigures \ref{fig:c}(a) and \ref{fig:c}(b) explicitly 
show how these couplings, initialized at zero in the UV, become nonzero as soon 
as the RG scale $k$ decreases. Their final numerical values in the IR limit, $k_{\mathrm{IR}} 
= 1\ \mathrm{MeV}$, are collected in the column ``Linear model'' of Table \ref{tab:LECs}.
\begin{table}[b!]
	\caption{\label{tab:LECs}Low-energy (derivative) couplings.\footnote{Values 
	are given at $k = 1\ \mathrm{MeV}$.}}
	\begin{ruledtabular}
		\begin{tabular}{lclc}
		\multicolumn{2}{c}{\textbf{Linear model}} & \multicolumn{2}{c}{\textbf{Nonlinear model}}\\[0.1cm]
		\colrule\\[-0.4cm]
		$\tilde{C}_{2}\ [1/f_{\pi}^{2}]\times 10$ & $-0.88$ & \multicolumn{2}{c}{$\ldots$} \\
		$\tilde{Z}_{2}\ [1/f_{\pi}^{2}]\times 10$ & $-2.30$ & 
		$\tilde{\mathcal{Z}}_{2}\ [1/f_{\pi}^{2}]\times 10$ & $-2.50$ \\
		$\tilde{C}_{3}\ [1/f_{\pi}^{4}]\times 10^{2}$ & $2.88$  & 
		$\tilde{\mathcal{C}}_{3}\ [1/f_{\pi}^{4}]\times 10^{2}$ & $-4.20$ \\
		$\tilde{C}_{4}\ [1/f_{\pi}^{4}]\times 10^{2}$ & $1.27$  & 
		$\tilde{\mathcal{C}}_{4}\ [1/f_{\pi}^{4}]\times 10^{2}$ & $1.27$ \\
		$\tilde{C}_{5}\ [1/f_{\pi}^{4}]\times 10^{2}$ & $4.69$  & 
		$\tilde{\mathcal{C}}_{5}\ [1/f_{\pi}^{4}]\times 10^{2}$ & $-2.41$ \\
		$\tilde{C}_{6}\ [1/f_{\pi}^{4}]\times 10^{2}$ & $-2.35$ & 
		$\tilde{\mathcal{C}}_{6}\ [1/f_{\pi}^{4}]\times 10^{2}$ & $1.21$ \\
		$\tilde{C}_{7}\ [1/f_{\pi}^{4}]\times 10^{2}$ & $0.02$  & \multicolumn{2}{c}{$\ldots$} \\
		$\tilde{C}_{8}\ [1/f_{\pi}^{4}]\times 10^{2}$ & $1.14$  & 
		$\tilde{\mathcal{C}}_{8}\ [1/f_{\pi}^{4}]\times 10^{2}$ & $-0.60$
		\end{tabular}
	\end{ruledtabular}
\end{table}

Since the $\mathcal{O}(\partial^{2})$ couplings of the nonlinear model do not depend 
on the corresponding couplings of the linear QMM, we discuss the $\mathcal{O}(\partial^{4})$
couplings first. Figure \ref{fig:c}(a) reveals that the main contribution to the couplings 
$\tilde{C}_{i,k}$, $i = 3,\ldots, 8$, comes from fluctuations with energy-momentum scales of 
$k \simeq 50 - 200\ \mathrm{MeV}$, which is significantly below the scale of spontaneous chiral symmetry
breaking; $k_{\text{SSB}}\simeq 300\ \mathrm{MeV}$. We thus conclude that these couplings are
well captured by the low-energy dynamics of the QMM. Also, given their slow running
above the scale of chiral symmetry breaking, the initial value of zero seems to be a very 
reasonable approximation for these couplings.

Although of less relevance for the low-energy couplings of the nonlinear model, it is still
interesting to investigate also the $\mathcal{O}(\partial^{2})$ couplings of the linear QMM.
As already observed in the preceding investigation \cite{Eser:2018jqo}, these couplings experience
a rapid initial change at RG scales close to the UV cutoff. This rapid initial change of the couplings 
of order $\mathcal{O}(\partial^{2})$ is an indication that a hadronic cutoff scale of around $500\ \mathrm{MeV}$ 
is actually too low for a precise determination of these couplings.
However, the fixed point-like scale evolution of $\tilde{C}_{2,k}$ and $\tilde{Z}_{2,k}$ at intermediate RG scales 
suggests that their IR values do not drastically dependent on the choice of $\Lambda$.
In future investigations, the dynamical-hadronization approach 
\cite{Gies:2001nw, Gies:2002hq, Pawlowski:2005xe, Floerchinger:2009uf} will allow for a smooth transition 
between the fundamental interactions and the bosonic operators presented in this study and, in turn,
for a computation of these couplings directly from quark and gluon fluctuations at QCD scales, 
cf.\ also Refs.\ \cite{Braun:2014ata, Mitter:2014wpa, Cyrol:2017ewj}.

\subsection{Nonlinear model: Low-energy couplings}
\label{sec:LECs}

As already argued, the computation of the low-energy limit of the QMM requires the 
integration of all nonpionic fields. Nevertheless, we present the nontrivial $\mathcal{O}(\partial^{4})$
low-energy couplings on all RG scales.

\begin{figure*}[t!]
	\centering
		\includegraphics{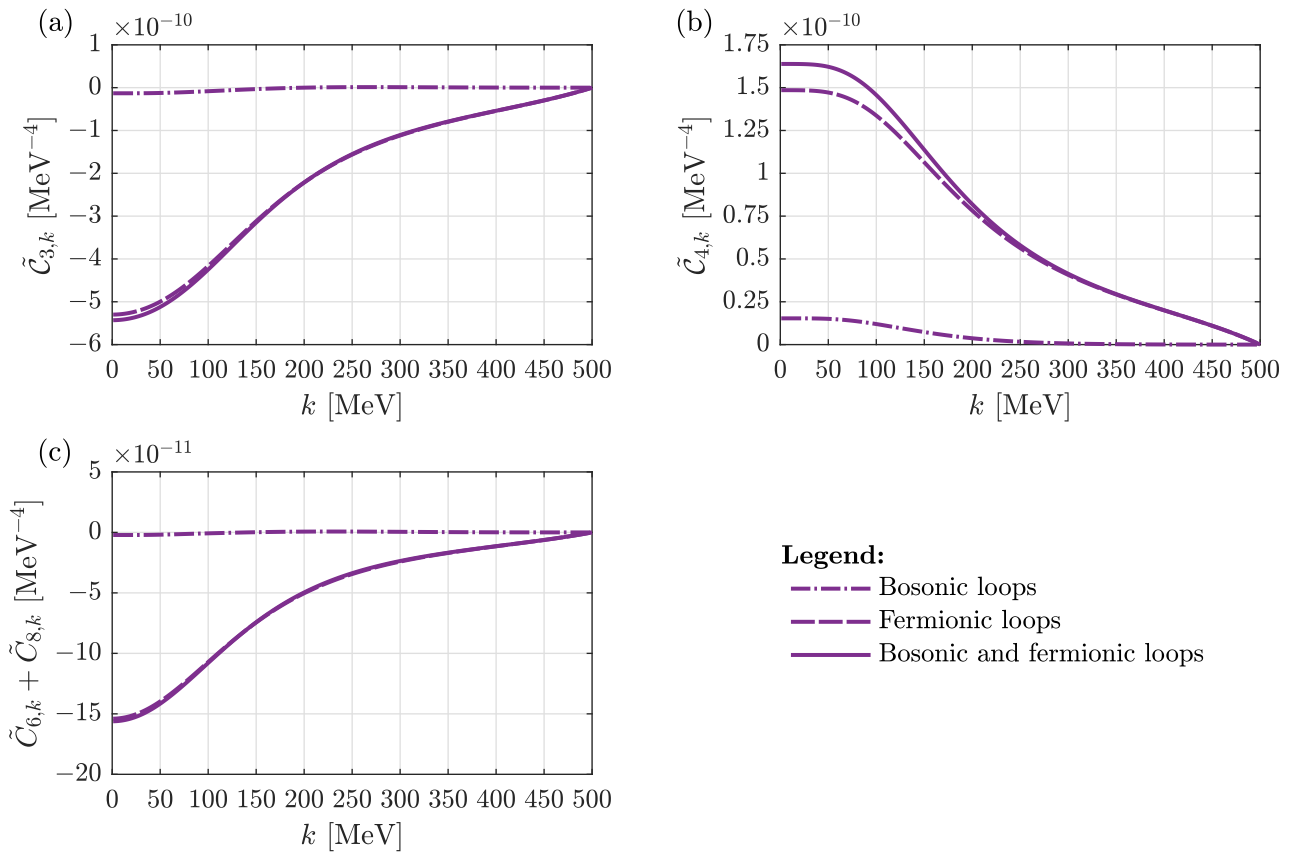}
	\caption{Low-energy couplings of the nonlinear model. (a) and (b)
	Scale evolution of the renormalized couplings
	$\tilde{\mathcal{C}}_{3,k}$ and $\tilde{\mathcal{C}}_{4,k}$, respectively. 
	(c) Scale evolution of $\tilde{C}_{6,k} + \tilde{C}_{8,k}$ representing 
	the renormalized couplings $\tilde{\mathcal{C}}_{5,k}$,
	$\tilde{\mathcal{C}}_{6,k}$, and $\tilde{\mathcal{C}}_{8,k}$. 
	The evolution of the couplings (solid lines) is decomposed into 
	fermionic (dashed lines) and bosonic contributions (dash-dotted lines); 
	$k_{\mathrm{IR}} = 1\ \mathrm{MeV}$.}
	\label{fig:cnonlinear}
\end{figure*}

Figure \ref{fig:cnonlinear} shows the results of
applying (\ref{eq:LECrelations}) at every RG scale. Additionally, we have decomposed 
the contributions to these couplings into quark and meson loops, respectively.
Despite the fact that the main contribution to the linear $\mathcal{O}(\partial^{4})$
couplings, cf.\ Fig.\ \ref{fig:c}(a), stems from fluctuations below the scale 
of chiral symmetry breaking, these couplings are almost exclusively determined by the quark 
fluctuations. The mesonic fluctuations and, in particular, the pionic fluctuations contribute 
noticeably to the total couplings only at energies below $200\ \mathrm{MeV}$. However,
also in this energy regime, the main contribution is due to quark fluctuations. This is 
in accordance with functional QCD calculations 
\cite{Braun:2014ata, Mitter:2014wpa, Cyrol:2017ewj} of other
low-energy couplings in the linear realization.

The most astonishing observation in Figs.\ \ref{fig:cnonlinear}(a), \ref{fig:cnonlinear}(b),
and \ref{fig:cnonlinear}(c) is that the loop contributions from the quark degrees of 
freedom are only fully integrated out at scales below $50 - 100\ \mathrm{MeV}$. This 
is almost certainly also the case in full QCD, since functional QCD calculations 
\cite{Braun:2014ata, Mitter:2014wpa, Cyrol:2017ewj} exhibit an even stronger dominance of 
qualitatively similar quark fluctuations. Therefore, we conclude that the low-energy couplings
can only be determined and defined from QCD below these scales, at least in the used 
renormalization scheme. Furthermore, this also leads to a natural cutoff scale for theories that
are exclusively based on pion fluctuations.

Lastly, a comparison with the according values from ChPT would require 
compatible renormalization schemes and comparable renormalization scales.
It is also clear that the computed low-energy couplings presented in this work
do not yet include the effect of resonances, especially, the light scalar
and vector channels. 

The numerical IR values of the higher-derivative couplings in the nonlinearly 
realized effective pion action are listed in the last column (``Nonlinear model'') 
of Table \ref{tab:LECs}. The couplings that vanish in the nonlinear picture are 
denoted by three dots. The pion mass amounts to $\tilde{\mathcal{M}}_{\Pi,k_{\mathrm{IR}}} = 
138.5\ \mathrm{MeV}$. The momentum-in\-de\-pen\-dent coupling $\tilde{\mathcal{C}}_{1,k_{\mathrm{IR}}}$ 
has a value of $0.27$.

\section{Summary and Outlook}
\label{sec:summary}

In this work, we studied the low-energy limit of the $O(4)$ QMM within the FRG approach 
by transforming the corresponding effective action into an effective pion theory. 
This corresponds to a transition from the QMM based on a linear realization
of the $O(4)$ symmetry to a model where the pions enter according to a nonlinear 
realization.

Our approach yields precise statements about the energy-momentum scale below
which the dynamics is dominated by pionic degrees of freedom. We find that 
for physical pion masses:
\begin{itemize}
	\item [(i)] The pion loop contributions to the low-energy couplings
    are strongly suppressed as compared to the quark loops.
    \item [(ii)] In our renormalization scheme, the scale for the decoupling of the quark fluctuations 
    and, in turn, the range of validity of the low-energy effective theory is roughly given by $50-100$ MeV. 
\end{itemize}    
Due to the qualitative similarity of the QMM dynamics to the low-energy limit of QCD, these statements most 
likely extrapolate to full QCD, which has to be checked explicitly in a future investigation.

Upcoming studies will also shed more light on the relation to ChPT, 
where this work can be understood as an extension of Refs.\ \cite{Jungnickel:1997yu, 
Divotgey:2016pst}. In particular, it would be very interesting if the computed low-energy 
couplings of the QMM and related effective theories, like the extended LSM \cite{Parganlija:2012fy}, 
are consistent with the low-energy limit of QCD (the latter is formalized by means of the LECs in ChPT). 
However, as stated above, this would require a meaningful choice of the renormalization scale 
in ChPT with regard to the physical relevance of pion fluctuations. Moreover, such a 
comparison implies a profound discussion of the effect of resonances on the low-energy
couplings within our approach. One also has to carefully evaluate how much this comparison 
would be distorted by the fact that our analysis is formulated in terms of the effective 
action, which has been generated from the flow of the linearly realized QMM.

Finally, and on a more technical level, the presented work is an extension of our previous 
exploratory study \cite{Eser:2018jqo} of higher-derivative pion self-interactions within the 
FRG formalism. The used truncation was improved by including higher-derivative $\sigma\pi$ 
as well as $\sigma$ self-interactions in a completely $O(4)$-symmetric way. Additionally, 
we took also the flow of the Yukawa coupling into account.
All higher-derivative couplings were dynamically generated from the FRG flow, which was
initialized at a hadronic scale of $500\ \mathrm{MeV}$. The large corrections of the 
interactions of order $\mathcal{O}(\partial^{2})$ right after the initialization suggests
a determination of these couplings from QCD scales in future investigations. Such an 
analysis could be carried out using the dynamical-hadronization procedure \cite{Braun:2014ata, 
Mitter:2014wpa, Cyrol:2017ewj}, which consistently describes mesonic degrees of freedom 
as quark-antiquark bound states.

\begin{acknowledgments}
The authors thank M.\ Birse, J.-P.\ Blaizot, D.\ D.\ Dietrich, J.\ Goity, A.\ Koenigstein,
J.\ M.\ Pawlowski, R.\ D.\ Pisarski, J.\ Qiu, S.\ Rechenberger, F.\ Rennecke, D.\ H.\ Rischke, B.-J.\ Schaefer, 
L.\ von Smekal, and C.\ Weiss for valuable discussions. 
J.\ E.\ acknowledges funding by the German National Academic Foundation and HIC for FAIR.
M.\ M.\ is supported by the DFG grant MI 2240/1-1 
and the U.S.\ Department of Energy under contract de-sc0012704. 
\end{acknowledgments}

\appendix

\section{Transformation properties}
\label{sec:trafoprops}

This section is intended to briefly review the transformation properties 
of the coset representative and the coefficients of the Maurer-Cartan form. 
It should be noted that $\Sigma(\zeta)$ is an element of $SO(4)$. Acting on 
the representative from the left by an arbitrary group element $g\in SO(4)$ clearly yields 
another group element $g\Sigma\left(\zeta\right)$. However, the latter object 
does not belong to the same left coset as $\Sigma\left(\zeta\right)$. Generally, 
the representative of this new coset takes the form $\Sigma\left(\zeta^{\prime}
\right)h\left(\zeta,g\right)$, such that \vspace{3pt}
\begin{equation}
	g\Sigma\left(\zeta\right) = \Sigma\left(\zeta^{\prime}\right)h\left(\zeta,g\right), \vspace{3pt}
	\label{eq:cosetreptrafo}
\end{equation} 
where $h\left(\zeta,g\right) \in SO(3)$ usually depends on the group element 
$g\in SO(4)$ and the pNGB fields. This implies that $h\left(\zeta,g\right)$ depends 
on spacetime and is therefore a local transformation. Then, Eq.\ (\ref{eq:cosetreptrafo}) 
automatically determines the transformation behavior of the pNGB fields,
\begin{equation}
	\zeta^{a} \xrightarrow{SO(4)} \left(\zeta^{a}\right)^{\prime} \equiv f^{a}\left(\zeta,g\right), 
	\quad a = 1,2,3,
	\label{eq:ngbtrafo}
\end{equation}
where the functions $f^{a}$ depend on the choice of coordinates and, in general, turn out to be nonlinear.

Using Eq.\ (\ref{eq:cosetreptrafo}) as well as Eqs.\ (\ref{eq:MCform1}) and 
(\ref{eq:MCform2}), the transformation properties of the coefficients of the 
horizontal and vertical part of the Maurer-Cartan form can be derived as
\begin{IEEEeqnarray}{rCl}
	e^{a}_{\mu}\left(\zeta\right)x_{a} & \xrightarrow{SO(4)} & h\left(\zeta,g\right)e^{a}_{\mu}
	\left(\zeta\right)x_{a}h^{-1}\left(\zeta,g\right) , \\
	\omega^{i}_{\mu}\left(\zeta\right)s_{i} & \xrightarrow{SO(4)} & h\left(\zeta,g\right)
	\omega^{i}_{\mu}
	\left(\zeta\right)s_{i}h^{-1}\left(\zeta,g\right) \nonumber \\
	& & -i h\left(\zeta,g\right)\partial_{\mu}h^{-1}\left(\zeta,g\right) .
\end{IEEEeqnarray}

In order to illustrate the above formulas by means of an example, we 
consider the transformation properties of the field variables of the 
models discussed in Secs.\ \ref{sec:LSM} and \ref{sec:effpion}. To this end, 
we start with the Lie algebra $\mathfrak{so}(4)$ and choose its basis according to
\begin{IEEEeqnarray}{rCl}
	\left[ s_{i}, s_{j}\right] & = & i\epsilon_{ij}^{\; \; \; k}s_{k}, \nonumber\\
	\left[ s_{i}, x_{b}\right] & = & i\epsilon_{ib}^{\; \; \; c}x_{c}, \nonumber\\
	\left[ x_{a}, x_{b}\right] & = & i\epsilon_{ab}^{\; \; \; k}s_{k},
\end{IEEEeqnarray}
where the indices $i,j,k=1,2,3$ label the generators $s_{i}$ of the 
unbroken $SO(3)$ subgroup and where $a,b,c=1,2,3$ label the coset 
generators $x_{a}$. An explicit realization of this basis is given by
\begin{IEEEeqnarray}{rCl}
	s_{1} & = & \begin{pmatrix} 0 & 0 & 0 & 0 \\ 0 & 0 & -i & 0 \\ 0 & i & 0 & 0 \\ 0 & 0 & 0 & 
	0\end{pmatrix}, \quad x_{1} = \begin{pmatrix} 0 & 0 & 0 & -i \\ 0 & 0 & 0 & 0 \\ 0 & 0 & 0 & 0 
	\\ i & 0 & 0 & 0\end{pmatrix}, \nonumber\\
	s_{2} & = & \begin{pmatrix} 0 & 0 & i & 0 \\ 0 & 0 & 0 & 0 \\ -i & 0 & 0 & 0 \\ 0 & 0 & 0 & 
	0\end{pmatrix}, \quad x_{2} = \begin{pmatrix} 0 & 0 & 0 & 0 \\ 0 & 0 & 0 & -i \\ 0 & 0 & 0 & 0 
	\\ 0 & i & 0 & 0\end{pmatrix}, \nonumber\\
	s_{3} & = & \begin{pmatrix} 0 & -i & 0 & 0 \\ i & 0 & 0 & 0 \\ 0 & 0 & 0 & 0 \\ 0 & 0 & 0 & 
	0\end{pmatrix}, \quad x_{3} = \begin{pmatrix} 0 & 0 & 0 & 0 \\ 0 & 0 & 0 & 0 \\ 0 & 0 & 0 & -i 
	\\ 0 & 0 & i & 0\end{pmatrix}.
\end{IEEEeqnarray}
The normalization of the above generators is chosen such that 
\begin{IEEEeqnarray}{rCl}
	\mathrm{tr}\left(s_{i}s_{j}\right) & = & 2\delta_{ij}, \nonumber\\
	\mathrm{tr}\left(x_{a}s_{b}\right) & = & 2\delta_{ab}, \nonumber\\
	\mathrm{tr}\left(s_{i}x_{a}\right) & = & 0,
\end{IEEEeqnarray}
for all $i,j=1,2,3$ and $a,b=1,2,3$.

Now, we consider an infinitesimal $SO(4)$ transformation
\begin{equation}
	O \simeq \mathbbmss{1}_{4} - i\alpha^{i}s_{i} - i\beta^{a}x_{a},
\end{equation}
where $\alpha^{i}$ and $\beta^{a}$ denote the group parameters. Using the isomorphism 
$SO(4)\cong SU(2)\times SU(2)$ as well as Eq.\ (\ref{eq:phitrafo}), one obtains
\begin{IEEEeqnarray}{rCl}
	\pi^{a} & \xrightarrow{SU(2)_{V}} & \pi^{a} + \epsilon_{jc}^{\;\;\;a}\alpha^{j}\pi^{c}, \quad 
	\sigma \xrightarrow{SU(2)_{V}} \sigma, \\
	\pi^{a} & \xrightarrow{SU(2)_{A}} & \pi^{a} - \beta^{a}\sigma, \quad \sigma \xrightarrow{SU(2)_{A}} 
	\sigma + \beta_{a}\pi^{a}, \quad
\end{IEEEeqnarray}
which is the usual transformation behavior of the $\sigma$ field and the pions. 
From these results and by using Eq.\ (\ref{eq:coords}), it is possible 
to derive the transformation behavior of the nonlinear fields. For the 
radial field $\theta$, one obtains
\begin{equation}
	\theta \xrightarrow{SU(2)_{V}} \theta, \qquad\qquad \theta \xrightarrow{SU(2)_{A}} \theta,
\end{equation}
while the pNGB fields transform as
\begin{IEEEeqnarray}{rCl}
	\zeta^{a} & \xrightarrow{SU(2)_{V}} & \zeta^{a} + \epsilon_{jc}^{\;\;\;a}\alpha^{j}\zeta^{c}, \\
	\zeta^{a} & \xrightarrow{SU(2)_{A}} & \zeta^{a} - \frac{\beta^{a}}{2}\left(1 - \zeta_{b}\zeta^{b}
	\right) - \zeta^{a}\left(\beta_{b}\zeta^{b}\right),
\end{IEEEeqnarray}
cf.\ Eq.\ (\ref{eq:ngbtrafo}). It is a general result that the pNGB fields transform in a linear 
representation with respect to the unbroken subgroup, whereas for 
coset transformations their transformation behavior becomes 
nonlinear.

\section{Flow equations}
\label{sec:floweqns}

The regulators $R_{k}$ in the Wetterich equation (\ref{eq:Wetterich}) 
effectively separate the low-energy modes in momentum 
space from the integration process by introducing
an additional mass term. As a function of the RG scale $k$ 
and the momentum $q$, they fulfill the limits
\begin{IEEEeqnarray}{rCl}
	R_{k} \rightarrow 0 & \ \text{for}\ & k \rightarrow 0, \\
	R_{k} \rightarrow \infty & \ \text{for}\ & k \rightarrow 
	\Lambda \rightarrow \infty , \\
	R_{k} > 0 & \ \text{for}\ & q \rightarrow 0, \\
	R_{k} \rightarrow 0 & \ \text{for}\ & q \rightarrow \infty .
\end{IEEEeqnarray}

The explicit expressions for the bosonic and fermionic regulators used in this work read
\begin{IEEEeqnarray}{rCl}
	R_{k}^{\sigma}\left(q^{2}\right) & = & Z_{k}^{\sigma} 
	q^{2}\, r \left(\frac{q^{2}}{k^{2}}\right), \label{eq:reg1}\\
	R_{k}^{\pi}\left(q^{2}\right) & = & Z_{k}^{\pi} q^{2}\, 
	r \left(\frac{q^{2}}{k^{2}}\right), \label{eq:reg2}\\
	R_{k}^{\psi}\left(q\right) & = & -iZ_{k}^{\psi} 
	\gamma_{\mu} q_{\mu}\, r \left(\frac{q^{2}}{k^{2}}\right), 
	\label{eq:reg3}
\end{IEEEeqnarray}
where we employ an exponential shape function $r$ of the generic form\ 
\cite{Wetterich:1989xg, Wetterich:1992yh, Litim:2001dt, Nandori:2012tc, 
Pawlowski:2015mlf}
\begin{equation}
	r(x) = \frac{x^{m-1}}{\exp\left(x^{m}\right) - 1},
\end{equation}
with $m = 1$ for both bosonic and fermionic degrees of freedom.

\begin{widetext}
The flow equations for the (unrenormalized) scale-dependent quantities corresponding to
truncation (\ref{eq:truncation}) are derived from the Wetterich equation 
(\ref{eq:Wetterich}) as follows [partly using a diagrammatic representation 
analogously to Eq.\ (\ref{eq:rep}); $\mathcal{V}$ denotes the infinite spacetime volume]:
\begin{IEEEeqnarray}{rCl}
	\partial_{k}U_{k} & = & \mathcal{V}^{-1} \partial_{k}\Gamma_{k}
	 = \mathcal{V}^{-1} \biggg( 
	\frac{1}{2} \! \! \vcenter{\hbox{
	\begin{pspicture}[showgrid=false](1.5,2.4)
		\psarc[linewidth=0.02,linestyle=dashed,dash=2pt 1pt,linecolor=blue](0.75,1.2){0.6}{115}{65}
		\pscircle[linewidth=0.03,fillstyle=solid,fillcolor=NavyBlue](0.75,1.8){0.25}
		\psline[linewidth=0.03](0.75,1.8)(0.92,1.97)
		\psline[linewidth=0.03](0.75,1.8)(0.58,1.97)
		\psline[linewidth=0.03](0.75,1.8)(0.58,1.63)
		\psline[linewidth=0.03](0.75,1.8)(0.92,1.63)
		\rput[b]{*0}(0.75,0.35){$\sigma$}
	\end{pspicture}
	}} \! \! \!
	+ \frac{1}{2} \! \!
	\vcenter{\hbox{
	\begin{pspicture}(1.5,2.4)
		\psarc[linewidth=0.02,linestyle=dashed,dash=2pt 1pt,linecolor=red](0.75,1.2){0.6}{115}{65}
		\pscircle[linewidth=0.03,fillstyle=solid,fillcolor=RedOrange](0.75,1.8){0.25}
		\psline[linewidth=0.03](0.75,1.8)(0.92,1.97)
		\psline[linewidth=0.03](0.75,1.8)(0.58,1.97)
		\psline[linewidth=0.03](0.75,1.8)(0.58,1.63)
		\psline[linewidth=0.03](0.75,1.8)(0.92,1.63)
		\rput[b]{*0}(0.75,0.35){$\pi$}
	\end{pspicture}
	}} \! \! \!
	- \! \! \!
	\vcenter{\hbox{
	\begin{pspicture}(1.5,2.4)
		\psarc[linewidth=0.02,arrowsize=2pt 3,arrowinset=0]{->}(0.75,1.2){0.6}{115}{189}
		\psarc[linewidth=0.02](0.75,1.2){0.6}{180}{65}
		\pscircle[linewidth=0.03,fillstyle=solid,fillcolor=gray](0.75,1.8){0.25}
		\psline[linewidth=0.03](0.75,1.8)(0.92,1.97)
		\psline[linewidth=0.03](0.75,1.8)(0.58,1.97)
		\psline[linewidth=0.03](0.75,1.8)(0.58,1.63)
		\psline[linewidth=0.03](0.75,1.8)(0.92,1.63)
		\rput[b]{*0}(0.75,0.2){$\psi$}
	\end{pspicture}
	}} \! \! \biggg), \\
	\partial_{k}y_{k} & = & \frac{1}{2} \mathcal{V}^{-1}
	\frac{1}{\sigma}\tr_{\gamma} \left[\frac{\delta}{\delta \bar{\psi}(0)}\, \partial_{k}\Gamma_{k}\,
	\frac{\overleftarrow{\delta}}{\delta\psi (0)}\right] \nonumber\\
	& = & \frac{1}{2} \mathcal{V}^{-1} \frac{1}{\sigma}
	\tr_{\gamma} \biggg(
	\frac{1}{2} \! \! \vcenter{\hbox{
	\begin{pspicture}(3.0,2.0)
		\psarc[linewidth=0.02,linestyle=dashed,dash=2pt 1pt,linecolor=blue](1.5,1.0){0.6}{20}{65}
		\psarc[linewidth=0.02,linestyle=dashed,dash=2pt 1pt,linecolor=blue](1.5,1.0){0.6}{115}{160}
		\psarc[linewidth=0.02,arrowsize=2pt 3,arrowinset=0]{->}(1.5,1.0){0.6}{200}{279}
		\psarc[linewidth=0.02](1.5,1.0){0.6}{270}{340}
		\psline[linewidth=0.02,ArrowInside=->,ArrowInsidePos=0.55,
		arrowsize=2pt 3,arrowinset=0](0.1,1.0)(0.7,1.0)
		\psline[linewidth=0.02,ArrowInside=->,ArrowInsidePos=0.55,
		arrowsize=2pt 3,arrowinset=0](2.3,1.0)(2.9,1.0)
		\pscircle[linewidth=0.03,fillstyle=solid,fillcolor=NavyBlue](1.5,1.6){0.25}
		\psline[linewidth=0.03](1.5,1.6)(1.67,1.77)
		\psline[linewidth=0.03](1.5,1.6)(1.33,1.77)
		\psline[linewidth=0.03](1.5,1.6)(1.33,1.43)
		\psline[linewidth=0.03](1.5,1.6)(1.67,1.43)
		\pscircle[linewidth=0.03,fillstyle=solid,fillcolor=lightgray](0.9,1.0){0.20}
		\pscircle[linewidth=0.03,fillstyle=solid,fillcolor=lightgray](2.1,1.0){0.20}
		\rput[b]{*0}(2.1,1.4){$\sigma$}
		\rput[b]{*0}(0.9,1.4){$\sigma$}
		\rput[b]{*0}(0.4,0.6){$\psi$}
		\rput[b]{*0}(2.6,0.6){$\psi$}
		\rput[b]{*0}(1.5,0.0){$\psi$}
		\rput[b]{*0}(0.9,0.92){\scriptsize{$3$}}
		\rput[b]{*0}(2.1,0.92){\scriptsize{$3$}}
	\end{pspicture}
	}} \! \! \! 
	+ \frac{1}{2} \! \! \vcenter{\hbox{
	\begin{pspicture}(3.0,2.0)
		\psarc[linewidth=0.02,linestyle=dashed,dash=2pt 1pt,linecolor=red](1.5,1.0){0.6}{20}{65}
		\psarc[linewidth=0.02,linestyle=dashed,dash=2pt 1pt,linecolor=red](1.5,1.0){0.6}{115}{160}
		\psarc[linewidth=0.02,arrowsize=2pt 3,arrowinset=0]{->}(1.5,1.0){0.6}{200}{279}
		\psarc[linewidth=0.02](1.5,1.0){0.6}{270}{340}
		\psline[linewidth=0.02,ArrowInside=->,ArrowInsidePos=0.55,
		arrowsize=2pt 3,arrowinset=0](0.1,1.0)(0.7,1.0)
		\psline[linewidth=0.02,ArrowInside=->,ArrowInsidePos=0.55,
		arrowsize=2pt 3,arrowinset=0](2.3,1.0)(2.9,1.0)
		\pscircle[linewidth=0.03,fillstyle=solid,fillcolor=RedOrange](1.5,1.6){0.25}
		\psline[linewidth=0.03](1.5,1.6)(1.67,1.77)
		\psline[linewidth=0.03](1.5,1.6)(1.33,1.77)
		\psline[linewidth=0.03](1.5,1.6)(1.33,1.43)
		\psline[linewidth=0.03](1.5,1.6)(1.67,1.43)
		\pscircle[linewidth=0.03,fillstyle=solid,fillcolor=lightgray](0.9,1.0){0.20}
		\pscircle[linewidth=0.03,fillstyle=solid,fillcolor=lightgray](2.1,1.0){0.20}
		\rput[b]{*0}(2.1,1.4){$\pi$}
		\rput[b]{*0}(0.9,1.4){$\pi$}
		\rput[b]{*0}(0.4,0.6){$\psi$}
		\rput[b]{*0}(2.6,0.6){$\psi$}
		\rput[b]{*0}(1.5,0.0){$\psi$}
		\rput[b]{*0}(0.9,0.92){\scriptsize{$3$}}
		\rput[b]{*0}(2.1,0.92){\scriptsize{$3$}}
	\end{pspicture}
	}} \! \! \! \nonumber\\
	& & \qquad \qquad \qquad \ - \! \! \vcenter{\hbox{
	\begin{pspicture}(3.0,2.0)
		\psarc[linewidth=0.02,arrowsize=2pt 3,arrowinset=0]{->}(1.5,1.0){0.6}{20}{55}
		\psarc[linewidth=0.02](1.5,1.0){0.6}{46}{65}
		\psarc[linewidth=0.02,arrowsize=2pt 3,arrowinset=0]{->}(1.5,1.0){0.6}{115}{150}
		\psarc[linewidth=0.02](1.5,1.0){0.6}{141}{160}
		\psarc[linewidth=0.02,linestyle=dashed,dash=2pt 1pt,linecolor=blue](1.5,1.0){0.6}{200}{340}
		\psline[linewidth=0.02,ArrowInside=->,ArrowInsidePos=0.55,
		arrowsize=2pt 3,arrowinset=0](0.7,1.0)(0.1,1.0)
		\psline[linewidth=0.02,ArrowInside=->,ArrowInsidePos=0.55,
		arrowsize=2pt 3,arrowinset=0](2.9,1.0)(2.3,1.0)
		\pscircle[linewidth=0.03,fillstyle=solid,fillcolor=gray](1.5,1.6){0.25}
		\psline[linewidth=0.03](1.5,1.6)(1.67,1.77)
		\psline[linewidth=0.03](1.5,1.6)(1.33,1.77)
		\psline[linewidth=0.03](1.5,1.6)(1.33,1.43)
		\psline[linewidth=0.03](1.5,1.6)(1.67,1.43)
		\pscircle[linewidth=0.03,fillstyle=solid,fillcolor=lightgray](0.9,1.0){0.20}
		\pscircle[linewidth=0.03,fillstyle=solid,fillcolor=lightgray](2.1,1.0){0.20}
		\rput[b]{*0}(2.1,1.4){$\psi$}
		\rput[b]{*0}(0.9,1.4){$\psi$}
		\rput[b]{*0}(0.4,0.6){$\psi$}
		\rput[b]{*0}(2.6,0.6){$\psi$}
		\rput[b]{*0}(1.5,0.15){$\sigma$}
		\rput[b]{*0}(0.9,0.92){\scriptsize{$3$}}
		\rput[b]{*0}(2.1,0.92){\scriptsize{$3$}}
	\end{pspicture}
	}} \! \! \! 
	- \! \! \! \vcenter{\hbox{
	\begin{pspicture}(3.0,2.0)
		\psarc[linewidth=0.02,arrowsize=2pt 3,arrowinset=0]{->}(1.5,1.0){0.6}{20}{55}
		\psarc[linewidth=0.02](1.5,1.0){0.6}{46}{65}
		\psarc[linewidth=0.02,arrowsize=2pt 3,arrowinset=0]{->}(1.5,1.0){0.6}{115}{150}
		\psarc[linewidth=0.02](1.5,1.0){0.6}{141}{160}
		\psarc[linewidth=0.02,linestyle=dashed,dash=2pt 1pt,linecolor=red](1.5,1.0){0.6}{200}{340}
		\psline[linewidth=0.02,ArrowInside=->,ArrowInsidePos=0.55,
		arrowsize=2pt 3,arrowinset=0](0.7,1.0)(0.1,1.0)
		\psline[linewidth=0.02,ArrowInside=->,ArrowInsidePos=0.55,
		arrowsize=2pt 3,arrowinset=0](2.9,1.0)(2.3,1.0)
		\pscircle[linewidth=0.03,fillstyle=solid,fillcolor=gray](1.5,1.6){0.25}
		\psline[linewidth=0.03](1.5,1.6)(1.67,1.77)
		\psline[linewidth=0.03](1.5,1.6)(1.33,1.77)
		\psline[linewidth=0.03](1.5,1.6)(1.33,1.43)
		\psline[linewidth=0.03](1.5,1.6)(1.67,1.43)
		\pscircle[linewidth=0.03,fillstyle=solid,fillcolor=lightgray](0.9,1.0){0.20}
		\pscircle[linewidth=0.03,fillstyle=solid,fillcolor=lightgray](2.1,1.0){0.20}
		\rput[b]{*0}(2.1,1.4){$\psi$}
		\rput[b]{*0}(0.9,1.4){$\psi$}
		\rput[b]{*0}(0.4,0.6){$\psi$}
		\rput[b]{*0}(2.6,0.6){$\psi$}
		\rput[b]{*0}(1.5,0.15){$\pi$}
		\rput[b]{*0}(0.9,0.92){\scriptsize{$3$}}
		\rput[b]{*0}(2.1,0.92){\scriptsize{$3$}}
	\end{pspicture}
	}} \! \! \biggg) ,\label{eq:yc} \\
	\partial_{k}Z_{k}^{\sigma} & = & \mathcal{V}^{-1} 
	\left.\frac{\mathrm{d}}{\mathrm{d}p^2}\right|_{p^{2}=0}
	\frac{\delta^{2}\partial_{k}\Gamma_{k}}{\delta\sigma(-p)\delta\sigma(p)} \nonumber\\
	& = & \mathcal{V}^{-1} \left.\frac{\mathrm{d}}{\mathrm{d}p^2}\right|_{p^{2}=0}\biggg(
	\frac{1}{2} \! \! \vcenter{\hbox{
	\begin{pspicture}[showgrid=false](3.0,2.0)
		\psarc[linewidth=0.02,linestyle=dashed,dash=2pt 1pt,linecolor=blue](1.5,1.0){0.6}{20}{65}
		\psarc[linewidth=0.02,linestyle=dashed,dash=2pt 1pt,linecolor=blue](1.5,1.0){0.6}{115}{160}
		\psarc[linewidth=0.02,linestyle=dashed,dash=2pt 1pt,linecolor=blue](1.5,1.0){0.6}{200}{340}
		\psline[linewidth=0.02,linestyle=dashed,dash=2pt 1pt,linecolor=blue](0.1,1.0)(0.7,1.0)
		\psline[linewidth=0.02,linestyle=dashed,dash=2pt 1pt,linecolor=blue](2.3,1.0)(2.9,1.0)
		\pscircle[linewidth=0.03,fillstyle=solid,fillcolor=NavyBlue](1.5,1.6){0.25}
		\psline[linewidth=0.03](1.5,1.6)(1.67,1.77)
		\psline[linewidth=0.03](1.5,1.6)(1.33,1.77)
		\psline[linewidth=0.03](1.5,1.6)(1.33,1.43)
		\psline[linewidth=0.03](1.5,1.6)(1.67,1.43)
		\pscircle[linewidth=0.03,fillstyle=solid,fillcolor=lightgray](0.9,1.0){0.20}
		\pscircle[linewidth=0.03,fillstyle=solid,fillcolor=lightgray](2.1,1.0){0.20}
		\rput[b]{*0}(2.1,1.4){$\sigma$}
		\rput[b]{*0}(0.9,1.4){$\sigma$}
		\rput[b]{*0}(0.4,0.7){$\sigma$}
		\rput[b]{*0}(2.6,0.7){$\sigma$}
		\rput[b]{*0}(1.5,0.15){$\sigma$}
		\rput[b]{*0}(0.9,0.92){\scriptsize{$3$}}
		\rput[b]{*0}(2.1,0.92){\scriptsize{$3$}}
	\end{pspicture}
	}} \! \! \! 
	 + \frac{1}{2} \! \! \vcenter{\hbox{
	\begin{pspicture}(3.0,2.0)
		\psarc[linewidth=0.02,linestyle=dashed,dash=2pt 1pt,linecolor=red](1.5,1.0){0.6}{20}{65}
		\psarc[linewidth=0.02,linestyle=dashed,dash=2pt 1pt,linecolor=red](1.5,1.0){0.6}{115}{160}
		\psarc[linewidth=0.02,linestyle=dashed,dash=2pt 1pt,linecolor=red](1.5,1.0){0.6}{200}{340}
		\psline[linewidth=0.02,linestyle=dashed,dash=2pt 1pt,linecolor=blue](0.1,1.0)(0.7,1.0)
		\psline[linewidth=0.02,linestyle=dashed,dash=2pt 1pt,linecolor=blue](2.3,1.0)(2.9,1.0)
		\pscircle[linewidth=0.03,fillstyle=solid,fillcolor=RedOrange](1.5,1.6){0.25}
		\psline[linewidth=0.03](1.5,1.6)(1.67,1.77)
		\psline[linewidth=0.03](1.5,1.6)(1.33,1.77)
		\psline[linewidth=0.03](1.5,1.6)(1.33,1.43)
		\psline[linewidth=0.03](1.5,1.6)(1.67,1.43)
		\pscircle[linewidth=0.03,fillstyle=solid,fillcolor=lightgray](0.9,1.0){0.20}
		\pscircle[linewidth=0.03,fillstyle=solid,fillcolor=lightgray](2.1,1.0){0.20}
		\rput[b]{*0}(2.1,1.4){$\pi$}
		\rput[b]{*0}(0.9,1.4){$\pi$}
		\rput[b]{*0}(0.4,0.7){$\sigma$}
		\rput[b]{*0}(2.6,0.7){$\sigma$}
		\rput[b]{*0}(1.5,0.15){$\pi$}
		\rput[b]{*0}(0.9,0.92){\scriptsize{$3$}}
		\rput[b]{*0}(2.1,0.92){\scriptsize{$3$}}
	\end{pspicture}
	}} \! \! \!  
	- \! \! \! \vcenter{\hbox{
	\begin{pspicture}(3.0,2.0)
		\psarc[linewidth=0.02,arrowsize=2pt 3,arrowinset=0]{->}(1.5,1.0){0.6}{20}{55}
		\psarc[linewidth=0.02](1.5,1.0){0.6}{46}{65}
		\psarc[linewidth=0.02,arrowsize=2pt 3,arrowinset=0]{->}(1.5,1.0){0.6}{115}{150}
		\psarc[linewidth=0.02](1.5,1.0){0.6}{141}{160}
		\psarc[linewidth=0.02,arrowsize=2pt 3,arrowinset=0]{->}(1.5,1.0){0.6}{200}{279}
		\psarc[linewidth=0.02](1.5,1.0){0.6}{270}{340}
		\psline[linewidth=0.02,linestyle=dashed,dash=2pt 1pt,linecolor=blue](0.1,1.0)(0.7,1.0)
		\psline[linewidth=0.02,linestyle=dashed,dash=2pt 1pt,linecolor=blue](2.3,1.0)(2.9,1.0)
		\pscircle[linewidth=0.03,fillstyle=solid,fillcolor=gray](1.5,1.6){0.25}
		\psline[linewidth=0.03](1.5,1.6)(1.67,1.77)
		\psline[linewidth=0.03](1.5,1.6)(1.33,1.77)
		\psline[linewidth=0.03](1.5,1.6)(1.33,1.43)
		\psline[linewidth=0.03](1.5,1.6)(1.67,1.43)
		\pscircle[linewidth=0.03,fillstyle=solid,fillcolor=lightgray](0.9,1.0){0.20}
		\pscircle[linewidth=0.03,fillstyle=solid,fillcolor=lightgray](2.1,1.0){0.20}
		\rput[b]{*0}(2.1,1.4){$\psi$}
		\rput[b]{*0}(0.9,1.4){$\psi$}
		\rput[b]{*0}(0.4,0.7){$\sigma$}
		\rput[b]{*0}(2.6,0.7){$\sigma$}
		\rput[b]{*0}(1.5,0.0){$\psi$}
		\rput[b]{*0}(0.9,0.92){\scriptsize{$3$}}
		\rput[b]{*0}(2.1,0.92){\scriptsize{$3$}}
	\end{pspicture}
	}} \! \! \nonumber\\ 
	& & \qquad \qquad \qquad \quad \ - \frac{1}{2} \! \!
	\vcenter{\hbox{
	\begin{pspicture}[showgrid=false](2.0,2.0)
		\psarc[linewidth=0.02,linestyle=dashed,dash=2pt 1pt,linecolor=blue](1.0,1.0){0.6}{290}{65}
		\psarc[linewidth=0.02,linestyle=dashed,dash=2pt 1pt,linecolor=blue](1.0,1.0){0.6}{115}{250}
		\psline[linewidth=0.02,linestyle=dashed,dash=2pt 1pt,linecolor=blue](0.35,0.1)(0.818408,0.316188)
		\psline[linewidth=0.02,linestyle=dashed,dash=2pt 1pt,linecolor=blue](1.181592,0.316188)(1.65,0.1)
		\pscircle[linewidth=0.03,fillstyle=solid,fillcolor=NavyBlue](1.0,1.6){0.25}
		\psline[linewidth=0.03](1.0,1.6)(1.17,1.77)
		\psline[linewidth=0.03](1.0,1.6)(0.83,1.77)
		\psline[linewidth=0.03](1.0,1.6)(0.83,1.43)
		\psline[linewidth=0.03](1.0,1.6)(1.17,1.43)
		\pscircle[linewidth=0.03,fillstyle=solid,fillcolor=lightgray](1.0,0.4){0.20}
		\rput[b]{*0}(0.985,0.32){\scriptsize{$4$}}
		\rput[b]{*0}(0.2,0.9){$\sigma$}
		\rput[b]{*0}(1.8,0.9){$\sigma$}
		\rput[b]{*0}(0.2,0.0){$\sigma$}
		\rput[b]{*0}(1.8,0.0){$\sigma$}
	\end{pspicture}
	}} \! \! \!
	- \frac{1}{2} \! \!
	\vcenter{\hbox{
	\begin{pspicture}[showgrid=false](2.0,2.0)
		\psarc[linewidth=0.02,linestyle=dashed,dash=2pt 1pt,linecolor=red](1.0,1.0){0.6}{290}{65}
		\psarc[linewidth=0.02,linestyle=dashed,dash=2pt 1pt,linecolor=red](1.0,1.0){0.6}{115}{250}
		\psline[linewidth=0.02,linestyle=dashed,dash=2pt 1pt,linecolor=blue](0.35,0.1)(0.818408,0.316188)
		\psline[linewidth=0.02,linestyle=dashed,dash=2pt 1pt,linecolor=blue](1.181592,0.316188)(1.65,0.1)
		\pscircle[linewidth=0.03,fillstyle=solid,fillcolor=RedOrange](1.0,1.6){0.25}
		\psline[linewidth=0.03](1.0,1.6)(1.17,1.77)
		\psline[linewidth=0.03](1.0,1.6)(0.83,1.77)
		\psline[linewidth=0.03](1.0,1.6)(0.83,1.43)
		\psline[linewidth=0.03](1.0,1.6)(1.17,1.43)
		\pscircle[linewidth=0.03,fillstyle=solid,fillcolor=lightgray](1.0,0.4){0.20}
		\rput[b]{*0}(0.985,0.32){\scriptsize{$4$}}
		\rput[b]{*0}(0.2,0.9){$\pi$}
		\rput[b]{*0}(1.8,0.9){$\pi$}
		\rput[b]{*0}(0.2,0.0){$\sigma$}
		\rput[b]{*0}(1.8,0.0){$\sigma$}
	\end{pspicture}
	}} \, \, \biggg) , \label{eq:zs} \\[0.2cm]
	\partial_{k}Z_{k}^{\pi} & = & \mathcal{V}^{-1} 
	\left.\frac{\mathrm{d}}{\mathrm{d}p^2}\right|_{p^{2}=0}
	\frac{\delta^{2}\partial_{k}\Gamma_{k}}{\delta\pi_{1}(-p)\delta\pi_{1}(p)} \nonumber\\
	& = & \mathcal{V}^{-1} \left.\frac{\mathrm{d}}{\mathrm{d}p^2}\right|_{p^{2}=0}\biggg(
	\frac{1}{2} \! \! \vcenter{\hbox{
	\begin{pspicture}(3.0,2.0)
		\psarc[linewidth=0.02,linestyle=dashed,dash=2pt 1pt,linecolor=blue](1.5,1.0){0.6}{20}{65}
		\psarc[linewidth=0.02,linestyle=dashed,dash=2pt 1pt,linecolor=blue](1.5,1.0){0.6}{115}{160}
		\psarc[linewidth=0.02,linestyle=dashed,dash=2pt 1pt,linecolor=red](1.5,1.0){0.6}{200}{340}
		\psline[linewidth=0.02,linestyle=dashed,dash=2pt 1pt,linecolor=red](0.1,1.0)(0.7,1.0)
		\psline[linewidth=0.02,linestyle=dashed,dash=2pt 1pt,linecolor=red](2.3,1.0)(2.9,1.0)
		\pscircle[linewidth=0.03,fillstyle=solid,fillcolor=NavyBlue](1.5,1.6){0.25}
		\psline[linewidth=0.03](1.5,1.6)(1.67,1.77)
		\psline[linewidth=0.03](1.5,1.6)(1.33,1.77)
		\psline[linewidth=0.03](1.5,1.6)(1.33,1.43)
		\psline[linewidth=0.03](1.5,1.6)(1.67,1.43)
		\pscircle[linewidth=0.03,fillstyle=solid,fillcolor=lightgray](0.9,1.0){0.20}
		\pscircle[linewidth=0.03,fillstyle=solid,fillcolor=lightgray](2.1,1.0){0.20}
		\rput[b]{*0}(2.1,1.4){$\sigma$}
		\rput[b]{*0}(0.9,1.4){$\sigma$}
		\rput[b]{*0}(0.4,0.7){$\pi$}
		\rput[b]{*0}(2.6,0.7){$\pi$}
		\rput[b]{*0}(1.5,0.15){$\pi$}
		\rput[b]{*0}(0.9,0.92){\scriptsize{$3$}}
		\rput[b]{*0}(2.1,0.92){\scriptsize{$3$}}
	\end{pspicture}
	}} \! \! \! 
	 + \frac{1}{2} \! \! \vcenter{\hbox{
	\begin{pspicture}(3.0,2.0)
		\psarc[linewidth=0.02,linestyle=dashed,dash=2pt 1pt,linecolor=red](1.5,1.0){0.6}{20}{65}
		\psarc[linewidth=0.02,linestyle=dashed,dash=2pt 1pt,linecolor=red](1.5,1.0){0.6}{115}{160}
		\psarc[linewidth=0.02,linestyle=dashed,dash=2pt 1pt,linecolor=blue](1.5,1.0){0.6}{200}{340}
		\psline[linewidth=0.02,linestyle=dashed,dash=2pt 1pt,linecolor=red](0.1,1.0)(0.7,1.0)
		\psline[linewidth=0.02,linestyle=dashed,dash=2pt 1pt,linecolor=red](2.3,1.0)(2.9,1.0)
		\pscircle[linewidth=0.03,fillstyle=solid,fillcolor=RedOrange](1.5,1.6){0.25}
		\psline[linewidth=0.03](1.5,1.6)(1.67,1.77)
		\psline[linewidth=0.03](1.5,1.6)(1.33,1.77)
		\psline[linewidth=0.03](1.5,1.6)(1.33,1.43)
		\psline[linewidth=0.03](1.5,1.6)(1.67,1.43)
		\pscircle[linewidth=0.03,fillstyle=solid,fillcolor=lightgray](0.9,1.0){0.20}
		\pscircle[linewidth=0.03,fillstyle=solid,fillcolor=lightgray](2.1,1.0){0.20}
		\rput[b]{*0}(2.1,1.4){$\pi$}
		\rput[b]{*0}(0.9,1.4){$\pi$}
		\rput[b]{*0}(0.4,0.7){$\pi$}
		\rput[b]{*0}(2.6,0.7){$\pi$}
		\rput[b]{*0}(1.5,0.15){$\sigma$}
		\rput[b]{*0}(0.9,0.92){\scriptsize{$3$}}
		\rput[b]{*0}(2.1,0.92){\scriptsize{$3$}}
	\end{pspicture}
	}} \! \! \!
	 - \! \! \! \vcenter{\hbox{
	\begin{pspicture}(3.0,2.0)
		\psarc[linewidth=0.02,arrowsize=2pt 3,arrowinset=0]{->}(1.5,1.0){0.6}{20}{55}
		\psarc[linewidth=0.02](1.5,1.0){0.6}{46}{65}
		\psarc[linewidth=0.02,arrowsize=2pt 3,arrowinset=0]{->}(1.5,1.0){0.6}{115}{150}
		\psarc[linewidth=0.02](1.5,1.0){0.6}{141}{160}
		\psarc[linewidth=0.02,arrowsize=2pt 3,arrowinset=0]{->}(1.5,1.0){0.6}{200}{279}
		\psarc[linewidth=0.02](1.5,1.0){0.6}{270}{340}
		\psline[linewidth=0.02,linestyle=dashed,dash=2pt 1pt,linecolor=red](0.1,1.0)(0.7,1.0)
		\psline[linewidth=0.02,linestyle=dashed,dash=2pt 1pt,linecolor=red](2.3,1.0)(2.9,1.0)
		\pscircle[linewidth=0.03,fillstyle=solid,fillcolor=gray](1.5,1.6){0.25}
		\psline[linewidth=0.03](1.5,1.6)(1.67,1.77)
		\psline[linewidth=0.03](1.5,1.6)(1.33,1.77)
		\psline[linewidth=0.03](1.5,1.6)(1.33,1.43)
		\psline[linewidth=0.03](1.5,1.6)(1.67,1.43)
		\pscircle[linewidth=0.03,fillstyle=solid,fillcolor=lightgray](0.9,1.0){0.20}
		\pscircle[linewidth=0.03,fillstyle=solid,fillcolor=lightgray](2.1,1.0){0.20}
		\rput[b]{*0}(2.1,1.4){$\psi$}
		\rput[b]{*0}(0.9,1.4){$\psi$}
		\rput[b]{*0}(0.4,0.7){$\pi$}
		\rput[b]{*0}(2.6,0.7){$\pi$}
		\rput[b]{*0}(1.5,0.0){$\psi$}
		\rput[b]{*0}(0.9,0.92){\scriptsize{$3$}}
		\rput[b]{*0}(2.1,0.92){\scriptsize{$3$}}
	\end{pspicture}
	}} \nonumber\\
	& & \qquad \qquad \qquad \quad \ - \frac{1}{2} \! \!
	\vcenter{\hbox{
	\begin{pspicture}[showgrid=false](2.0,2.0)
		\psarc[linewidth=0.02,linestyle=dashed,dash=2pt 1pt,linecolor=blue](1.0,1.0){0.6}{290}{65}
		\psarc[linewidth=0.02,linestyle=dashed,dash=2pt 1pt,linecolor=blue](1.0,1.0){0.6}{115}{250}
		\psline[linewidth=0.02,linestyle=dashed,dash=2pt 1pt,linecolor=red](0.35,0.1)(0.818408,0.316188)
		\psline[linewidth=0.02,linestyle=dashed,dash=2pt 1pt,linecolor=red](1.181592,0.316188)(1.65,0.1)
		\pscircle[linewidth=0.03,fillstyle=solid,fillcolor=NavyBlue](1.0,1.6){0.25}
		\psline[linewidth=0.03](1.0,1.6)(1.17,1.77)
		\psline[linewidth=0.03](1.0,1.6)(0.83,1.77)
		\psline[linewidth=0.03](1.0,1.6)(0.83,1.43)
		\psline[linewidth=0.03](1.0,1.6)(1.17,1.43)
		\pscircle[linewidth=0.03,fillstyle=solid,fillcolor=lightgray](1.0,0.4){0.20}
		\rput[b]{*0}(0.985,0.32){\scriptsize{$4$}}
		\rput[b]{*0}(0.2,0.9){$\sigma$}
		\rput[b]{*0}(1.8,0.9){$\sigma$}
		\rput[b]{*0}(0.2,0.0){$\pi$}
		\rput[b]{*0}(1.8,0.0){$\pi$}
	\end{pspicture}
	}} \! \! \!
	- \frac{1}{2} \! \!
	\vcenter{\hbox{
	\begin{pspicture}[showgrid=false](2.0,2.0)
		\psarc[linewidth=0.02,linestyle=dashed,dash=2pt 1pt,linecolor=red](1.0,1.0){0.6}{290}{65}
		\psarc[linewidth=0.02,linestyle=dashed,dash=2pt 1pt,linecolor=red](1.0,1.0){0.6}{115}{250}
		\psline[linewidth=0.02,linestyle=dashed,dash=2pt 1pt,linecolor=red](0.35,0.1)(0.818408,0.316188)
		\psline[linewidth=0.02,linestyle=dashed,dash=2pt 1pt,linecolor=red](1.181592,0.316188)(1.65,0.1)
		\pscircle[linewidth=0.03,fillstyle=solid,fillcolor=RedOrange](1.0,1.6){0.25}
		\psline[linewidth=0.03](1.0,1.6)(1.17,1.77)
		\psline[linewidth=0.03](1.0,1.6)(0.83,1.77)
		\psline[linewidth=0.03](1.0,1.6)(0.83,1.43)
		\psline[linewidth=0.03](1.0,1.6)(1.17,1.43)
		\pscircle[linewidth=0.03,fillstyle=solid,fillcolor=lightgray](1.0,0.4){0.20}
		\rput[b]{*0}(0.985,0.32){\scriptsize{$4$}}
		\rput[b]{*0}(0.2,0.9){$\pi$}
		\rput[b]{*0}(1.8,0.9){$\pi$}
		\rput[b]{*0}(0.2,0.0){$\pi$}
		\rput[b]{*0}(1.8,0.0){$\pi$}
	\end{pspicture}
	}} \, \, \biggg),\label{eq:zp} \\[0.2cm]
	\partial_{k}Z_{k}^{\psi} & = & \frac{i}{4} \mathcal{V}^{-1}
	\left.\frac{\mathrm{d}}{\mathrm{d}p^2}\right|_{p^{2}=0}
	\tr_{\gamma} \left[\frac{\delta}{\delta \bar{\psi}(p)}\, \partial_{k}\Gamma_{k}\,
	\frac{\overleftarrow{\delta}}{\delta\psi (p)}\ \gamma_{\mu}p_{\mu} \right] \nonumber\\
	& = & \frac{i}{4} \mathcal{V}^{-1}
	\left.\frac{\mathrm{d}}{\mathrm{d}p^2}\right|_{p^{2}=0}
	\tr_{\gamma} \biggg[\biggg(
	\frac{1}{2} \! \! \vcenter{\hbox{
	\begin{pspicture}(3.0,2.0)
		\psarc[linewidth=0.02,linestyle=dashed,dash=2pt 1pt,linecolor=blue](1.5,1.0){0.6}{20}{65}
		\psarc[linewidth=0.02,linestyle=dashed,dash=2pt 1pt,linecolor=blue](1.5,1.0){0.6}{115}{160}
		\psarc[linewidth=0.02,arrowsize=2pt 3,arrowinset=0]{->}(1.5,1.0){0.6}{200}{279}
		\psarc[linewidth=0.02](1.5,1.0){0.6}{270}{340}
		\psline[linewidth=0.02,ArrowInside=->,ArrowInsidePos=0.55,
		arrowsize=2pt 3,arrowinset=0](0.1,1.0)(0.7,1.0)
		\psline[linewidth=0.02,ArrowInside=->,ArrowInsidePos=0.55,
		arrowsize=2pt 3,arrowinset=0](2.3,1.0)(2.9,1.0)
		\pscircle[linewidth=0.03,fillstyle=solid,fillcolor=NavyBlue](1.5,1.6){0.25}
		\psline[linewidth=0.03](1.5,1.6)(1.67,1.77)
		\psline[linewidth=0.03](1.5,1.6)(1.33,1.77)
		\psline[linewidth=0.03](1.5,1.6)(1.33,1.43)
		\psline[linewidth=0.03](1.5,1.6)(1.67,1.43)
		\pscircle[linewidth=0.03,fillstyle=solid,fillcolor=lightgray](0.9,1.0){0.20}
		\pscircle[linewidth=0.03,fillstyle=solid,fillcolor=lightgray](2.1,1.0){0.20}
		\rput[b]{*0}(2.1,1.4){$\sigma$}
		\rput[b]{*0}(0.9,1.4){$\sigma$}
		\rput[b]{*0}(0.4,0.6){$\psi$}
		\rput[b]{*0}(2.6,0.6){$\psi$}
		\rput[b]{*0}(1.5,0.0){$\psi$}
		\rput[b]{*0}(0.9,0.92){\scriptsize{$3$}}
		\rput[b]{*0}(2.1,0.92){\scriptsize{$3$}}
	\end{pspicture}
	}} \! \! \! 
	+ \frac{1}{2} \! \! \vcenter{\hbox{
	\begin{pspicture}(3.0,2.0)
		\psarc[linewidth=0.02,linestyle=dashed,dash=2pt 1pt,linecolor=red](1.5,1.0){0.6}{20}{65}
		\psarc[linewidth=0.02,linestyle=dashed,dash=2pt 1pt,linecolor=red](1.5,1.0){0.6}{115}{160}
		\psarc[linewidth=0.02,arrowsize=2pt 3,arrowinset=0]{->}(1.5,1.0){0.6}{200}{279}
		\psarc[linewidth=0.02](1.5,1.0){0.6}{270}{340}
		\psline[linewidth=0.02,ArrowInside=->,ArrowInsidePos=0.55,
		arrowsize=2pt 3,arrowinset=0](0.1,1.0)(0.7,1.0)
		\psline[linewidth=0.02,ArrowInside=->,ArrowInsidePos=0.55,
		arrowsize=2pt 3,arrowinset=0](2.3,1.0)(2.9,1.0)
		\pscircle[linewidth=0.03,fillstyle=solid,fillcolor=RedOrange](1.5,1.6){0.25}
		\psline[linewidth=0.03](1.5,1.6)(1.67,1.77)
		\psline[linewidth=0.03](1.5,1.6)(1.33,1.77)
		\psline[linewidth=0.03](1.5,1.6)(1.33,1.43)
		\psline[linewidth=0.03](1.5,1.6)(1.67,1.43)
		\pscircle[linewidth=0.03,fillstyle=solid,fillcolor=lightgray](0.9,1.0){0.20}
		\pscircle[linewidth=0.03,fillstyle=solid,fillcolor=lightgray](2.1,1.0){0.20}
		\rput[b]{*0}(2.1,1.4){$\pi$}
		\rput[b]{*0}(0.9,1.4){$\pi$}
		\rput[b]{*0}(0.4,0.6){$\psi$}
		\rput[b]{*0}(2.6,0.6){$\psi$}
		\rput[b]{*0}(1.5,0.0){$\psi$}
		\rput[b]{*0}(0.9,0.92){\scriptsize{$3$}}
		\rput[b]{*0}(2.1,0.92){\scriptsize{$3$}}
	\end{pspicture}
	}} \! \! \! \nonumber\\
	& & \qquad \qquad \qquad \qquad \qquad \ - \! \! \vcenter{\hbox{
	\begin{pspicture}(3.0,2.0)
		\psarc[linewidth=0.02,arrowsize=2pt 3,arrowinset=0]{->}(1.5,1.0){0.6}{20}{55}
		\psarc[linewidth=0.02](1.5,1.0){0.6}{46}{65}
		\psarc[linewidth=0.02,arrowsize=2pt 3,arrowinset=0]{->}(1.5,1.0){0.6}{115}{150}
		\psarc[linewidth=0.02](1.5,1.0){0.6}{141}{160}
		\psarc[linewidth=0.02,linestyle=dashed,dash=2pt 1pt,linecolor=blue](1.5,1.0){0.6}{200}{340}
		\psline[linewidth=0.02,ArrowInside=->,ArrowInsidePos=0.55,
		arrowsize=2pt 3,arrowinset=0](0.7,1.0)(0.1,1.0)
		\psline[linewidth=0.02,ArrowInside=->,ArrowInsidePos=0.55,
		arrowsize=2pt 3,arrowinset=0](2.9,1.0)(2.3,1.0)
		\pscircle[linewidth=0.03,fillstyle=solid,fillcolor=gray](1.5,1.6){0.25}
		\psline[linewidth=0.03](1.5,1.6)(1.67,1.77)
		\psline[linewidth=0.03](1.5,1.6)(1.33,1.77)
		\psline[linewidth=0.03](1.5,1.6)(1.33,1.43)
		\psline[linewidth=0.03](1.5,1.6)(1.67,1.43)
		\pscircle[linewidth=0.03,fillstyle=solid,fillcolor=lightgray](0.9,1.0){0.20}
		\pscircle[linewidth=0.03,fillstyle=solid,fillcolor=lightgray](2.1,1.0){0.20}
		\rput[b]{*0}(2.1,1.4){$\psi$}
		\rput[b]{*0}(0.9,1.4){$\psi$}
		\rput[b]{*0}(0.4,0.6){$\psi$}
		\rput[b]{*0}(2.6,0.6){$\psi$}
		\rput[b]{*0}(1.5,0.15){$\sigma$}
		\rput[b]{*0}(0.9,0.92){\scriptsize{$3$}}
		\rput[b]{*0}(2.1,0.92){\scriptsize{$3$}}
	\end{pspicture}
	}} \! \! \! 
	- \! \! \! \vcenter{\hbox{
	\begin{pspicture}(3.0,2.0)
		\psarc[linewidth=0.02,arrowsize=2pt 3,arrowinset=0]{->}(1.5,1.0){0.6}{20}{55}
		\psarc[linewidth=0.02](1.5,1.0){0.6}{46}{65}
		\psarc[linewidth=0.02,arrowsize=2pt 3,arrowinset=0]{->}(1.5,1.0){0.6}{115}{150}
		\psarc[linewidth=0.02](1.5,1.0){0.6}{141}{160}
		\psarc[linewidth=0.02,linestyle=dashed,dash=2pt 1pt,linecolor=red](1.5,1.0){0.6}{200}{340}
		\psline[linewidth=0.02,ArrowInside=->,ArrowInsidePos=0.55,
		arrowsize=2pt 3,arrowinset=0](0.7,1.0)(0.1,1.0)
		\psline[linewidth=0.02,ArrowInside=->,ArrowInsidePos=0.55,
		arrowsize=2pt 3,arrowinset=0](2.9,1.0)(2.3,1.0)
		\pscircle[linewidth=0.03,fillstyle=solid,fillcolor=gray](1.5,1.6){0.25}
		\psline[linewidth=0.03](1.5,1.6)(1.67,1.77)
		\psline[linewidth=0.03](1.5,1.6)(1.33,1.77)
		\psline[linewidth=0.03](1.5,1.6)(1.33,1.43)
		\psline[linewidth=0.03](1.5,1.6)(1.67,1.43)
		\pscircle[linewidth=0.03,fillstyle=solid,fillcolor=lightgray](0.9,1.0){0.20}
		\pscircle[linewidth=0.03,fillstyle=solid,fillcolor=lightgray](2.1,1.0){0.20}
		\rput[b]{*0}(2.1,1.4){$\psi$}
		\rput[b]{*0}(0.9,1.4){$\psi$}
		\rput[b]{*0}(0.4,0.6){$\psi$}
		\rput[b]{*0}(2.6,0.6){$\psi$}
		\rput[b]{*0}(1.5,0.15){$\pi$}
		\rput[b]{*0}(0.9,0.92){\scriptsize{$3$}}
		\rput[b]{*0}(2.1,0.92){\scriptsize{$3$}}
	\end{pspicture}
	}} \! \! \biggg) \gamma_{\mu}p_{\mu} \biggg],\label{eq:zf} \\[0.2cm]
	\partial_{k}C_{2,k} & = & \frac{1}{2} \mathcal{V}^{-1} 
	\left.\frac{\mathrm{d}}{\mathrm{d}p^2}\right|_{p^{2}=0}
	\frac{\delta^{4}\partial_{k}\Gamma_{k}}{\delta\pi_{1}(p)
	\delta\pi_{2}(-p)\delta\pi_{1}(0)\delta\pi_{2}(0)} \nonumber\\
	& = & \frac{1}{2} \mathcal{V}^{-1} 
	\left.\frac{\mathrm{d}}{\mathrm{d}p^2}\right|_{p^{2}=0}\Biggg(
	- \frac{1}{2} \! \! \vcenter{\hbox{
	\begin{pspicture}[showgrid=false](3.0,2.0)
		\psarc[linewidth=0.02,linestyle=dashed,dash=2pt 1pt,linecolor=blue](1.5,1.0){0.6}{20}{65}
		\psarc[linewidth=0.02,linestyle=dashed,dash=2pt 1pt,linecolor=blue](1.5,1.0){0.6}{115}{160}
		\psarc[linewidth=0.02,linestyle=dashed,dash=2pt 1pt,linecolor=red](1.5,1.0){0.6}{290}{340}
		\psarc[linewidth=0.02,linestyle=dashed,dash=2pt 1pt,linecolor=red](1.5,1.0){0.6}{200}{250}
		\psline[linewidth=0.02,linestyle=dashed,dash=2pt 1pt,linecolor=red](0.85,0.1)(1.318408,0.316188)
		\psline[linewidth=0.02,linestyle=dashed,dash=2pt 1pt,linecolor=red](1.681592,0.316188)(2.15,0.1)
		\psline[linewidth=0.02,linestyle=dashed,dash=2pt 1pt,linecolor=red](0.1,1.0)(0.7,1.0)
		\psline[linewidth=0.02,linestyle=dashed,dash=2pt 1pt,linecolor=red](2.3,1.0)(2.9,1.0)
		\pscircle[linewidth=0.03,fillstyle=solid,fillcolor=NavyBlue](1.5,1.6){0.25}
		\psline[linewidth=0.03](1.5,1.6)(1.67,1.77)
		\psline[linewidth=0.03](1.5,1.6)(1.33,1.77)
		\psline[linewidth=0.03](1.5,1.6)(1.33,1.43)
		\psline[linewidth=0.03](1.5,1.6)(1.67,1.43)
		\pscircle[linewidth=0.03,fillstyle=solid,fillcolor=lightgray](1.5,0.4){0.20}
		\rput[b]{*0}(1.485,0.32){\scriptsize{$4$}}
		\pscircle[linewidth=0.03,fillstyle=solid,fillcolor=lightgray](0.9,1.0){0.20}
		\pscircle[linewidth=0.03,fillstyle=solid,fillcolor=lightgray](2.1,1.0){0.20}
		\rput[b]{*0}(0.9,0.92){\scriptsize{$3$}}
		\rput[b]{*0}(2.1,0.92){\scriptsize{$3$}}
		\rput[b]{*0}(0.4,0.7){$\pi$}
		\rput[b]{*0}(2.6,0.7){$\pi$}
		\rput[b]{*0}(0.7,0.0){$\pi$}
		\rput[b]{*0}(2.3,0.0){$\pi$}
		\rput[b]{*0}(2.1,1.4){$\sigma$}
		\rput[b]{*0}(0.9,1.4){$\sigma$}
		\rput[t]{*0}(2.1,0.55){$\pi$}
		\rput[t]{*0}(0.9,0.55){$\pi$}
	\end{pspicture}
	}} \! \! \! 
	- \frac{1}{2} \! \! \vcenter{\hbox{
	\begin{pspicture}(3.0,2.0)
		\psarc[linewidth=0.02,linestyle=dashed,dash=2pt 1pt,linecolor=red](1.5,1.0){0.6}{20}{65}
		\psarc[linewidth=0.02,linestyle=dashed,dash=2pt 1pt,linecolor=red](1.5,1.0){0.6}{115}{160}
		\psarc[linewidth=0.02,linestyle=dashed,dash=2pt 1pt,linecolor=blue](1.5,1.0){0.6}{290}{340}
		\psarc[linewidth=0.02,linestyle=dashed,dash=2pt 1pt,linecolor=blue](1.5,1.0){0.6}{200}{250}
		\psline[linewidth=0.02,linestyle=dashed,dash=2pt 1pt,linecolor=red](0.85,0.1)(1.318408,0.316188)
		\psline[linewidth=0.02,linestyle=dashed,dash=2pt 1pt,linecolor=red](1.681592,0.316188)(2.15,0.1)
		\psline[linewidth=0.02,linestyle=dashed,dash=2pt 1pt,linecolor=red](0.1,1.0)(0.7,1.0)
		\psline[linewidth=0.02,linestyle=dashed,dash=2pt 1pt,linecolor=red](2.3,1.0)(2.9,1.0)
		\pscircle[linewidth=0.03,fillstyle=solid,fillcolor=RedOrange](1.5,1.6){0.25}
		\psline[linewidth=0.03](1.5,1.6)(1.67,1.77)
		\psline[linewidth=0.03](1.5,1.6)(1.33,1.77)
		\psline[linewidth=0.03](1.5,1.6)(1.33,1.43)
		\psline[linewidth=0.03](1.5,1.6)(1.67,1.43)
		\pscircle[linewidth=0.03,fillstyle=solid,fillcolor=lightgray](1.5,0.4){0.20}
		\rput[b]{*0}(1.485,0.32){\scriptsize{$4$}}
		\pscircle[linewidth=0.03,fillstyle=solid,fillcolor=lightgray](0.9,1.0){0.20}
		\pscircle[linewidth=0.03,fillstyle=solid,fillcolor=lightgray](2.1,1.0){0.20}
		\rput[b]{*0}(0.9,0.92){\scriptsize{$3$}}
		\rput[b]{*0}(2.1,0.92){\scriptsize{$3$}}
		\rput[b]{*0}(0.4,0.7){$\pi$}
		\rput[b]{*0}(2.6,0.7){$\pi$}
		\rput[b]{*0}(0.7,0.0){$\pi$}
		\rput[b]{*0}(2.3,0.0){$\pi$}
		\rput[b]{*0}(2.1,1.4){$\pi$}
		\rput[b]{*0}(0.9,1.4){$\pi$}
		\rput[t]{*0}(2.1,0.55){$\sigma$}
		\rput[t]{*0}(0.9,0.55){$\sigma$}
	\end{pspicture}
	}} \vspace{-0.7cm} \nonumber\\
	& & \qquad \qquad \qquad \qquad \ + \ \frac{1}{2} \! \! \vcenter{\hbox{
	\begin{pspicture}[showgrid=false](2.6,2.0)
		\psarc[linewidth=0.02,linestyle=dashed,dash=2pt 1pt,linecolor=blue](1.3,1.0){0.6}{20}{65}
		\psarc[linewidth=0.02,linestyle=dashed,dash=2pt 1pt,linecolor=blue](1.3,1.0){0.6}{115}{160}
		\psarc[linewidth=0.02,linestyle=dashed,dash=2pt 1pt,linecolor=blue](1.3,1.0){0.6}{200}{340}
		\psline[linewidth=0.02,linestyle=dashed,dash=2pt 1pt,linecolor=red](1.98381,1.18159)(2.2,1.65)
		\psline[linewidth=0.02,linestyle=dashed,dash=2pt 1pt,linecolor=red](1.98381,0.818408)(2.2,0.35)
		\psline[linewidth=0.02,linestyle=dashed,dash=2pt 1pt,linecolor=red](0.616188,1.18159)(0.4,1.65)
		\psline[linewidth=0.02,linestyle=dashed,dash=2pt 1pt,linecolor=red](0.616188,0.818408)(0.4,0.35)
		\pscircle[linewidth=0.03,fillstyle=solid,fillcolor=NavyBlue](1.3,1.6){0.25}
		\psline[linewidth=0.03](1.3,1.6)(1.47,1.77)
		\psline[linewidth=0.03](1.3,1.6)(1.13,1.77)
		\psline[linewidth=0.03](1.3,1.6)(1.13,1.43)
		\psline[linewidth=0.03](1.3,1.6)(1.47,1.43)
		\pscircle[linewidth=0.03,fillstyle=solid,fillcolor=lightgray](0.7,1.0){0.20}
		\pscircle[linewidth=0.03,fillstyle=solid,fillcolor=lightgray](1.9,1.0){0.20}
		\rput[b]{*0}(0.685,0.92){\scriptsize{$4$}}
		\rput[b]{*0}(1.885,0.92){\scriptsize{$4$}}
		\rput[b]{*0}(0.2,0.2){$\pi$}
		\rput[b]{*0}(2.4,0.2){$\pi$}
		\rput[b]{*0}(0.2,1.7){$\pi$}
		\rput[b]{*0}(2.4,1.7){$\pi$}
		\rput[b]{*0}(1.9,1.4){$\sigma$}
		\rput[b]{*0}(0.7,1.4){$\sigma$}
		\rput[b]{*0}(1.3,0.15){$\sigma$}
	\end{pspicture}
	}} \! \! \!
	 + \frac{1}{2} \! \! \vcenter{\hbox{
	\begin{pspicture}(2.6,2.0)
		\psarc[linewidth=0.02,linestyle=dashed,dash=2pt 1pt,linecolor=red](1.3,1.0){0.6}{20}{65}
		\psarc[linewidth=0.02,linestyle=dashed,dash=2pt 1pt,linecolor=red](1.3,1.0){0.6}{115}{160}
		\psarc[linewidth=0.02,linestyle=dashed,dash=2pt 1pt,linecolor=red](1.3,1.0){0.6}{200}{340}
		\psline[linewidth=0.02,linestyle=dashed,dash=2pt 1pt,linecolor=red](1.98381,1.18159)(2.2,1.65)
		\psline[linewidth=0.02,linestyle=dashed,dash=2pt 1pt,linecolor=red](1.98381,0.818408)(2.2,0.35)
		\psline[linewidth=0.02,linestyle=dashed,dash=2pt 1pt,linecolor=red](0.616188,1.18159)(0.4,1.65)
		\psline[linewidth=0.02,linestyle=dashed,dash=2pt 1pt,linecolor=red](0.616188,0.818408)(0.4,0.35)
		\pscircle[linewidth=0.03,fillstyle=solid,fillcolor=RedOrange](1.3,1.6){0.25}
		\psline[linewidth=0.03](1.3,1.6)(1.47,1.77)
		\psline[linewidth=0.03](1.3,1.6)(1.13,1.77)
		\psline[linewidth=0.03](1.3,1.6)(1.13,1.43)
		\psline[linewidth=0.03](1.3,1.6)(1.47,1.43)
		\pscircle[linewidth=0.03,fillstyle=solid,fillcolor=lightgray](0.7,1.0){0.20}
		\pscircle[linewidth=0.03,fillstyle=solid,fillcolor=lightgray](1.9,1.0){0.20}
		\rput[b]{*0}(0.685,0.92){\scriptsize{$4$}}
		\rput[b]{*0}(1.885,0.92){\scriptsize{$4$}}
		\rput[b]{*0}(0.2,0.2){$\pi$}
		\rput[b]{*0}(2.4,0.2){$\pi$}
		\rput[b]{*0}(0.2,1.7){$\pi$}
		\rput[b]{*0}(2.4,1.7){$\pi$}
		\rput[b]{*0}(1.9,1.4){$\pi$}
		\rput[b]{*0}(0.7,1.4){$\pi$}
		\rput[b]{*0}(1.3,0.15){$\pi$}
	\end{pspicture}
	}} \! \! \!
	 - \frac{1}{2} \! \! \vcenter{\hbox{
	\begin{pspicture}[showgrid=false](3.0,3.2)
		\psarc[linewidth=0.02,linestyle=dashed,dash=2pt 1pt,linecolor=blue](1.5,1.6){0.6}{20}{65}
		\psarc[linewidth=0.02,linestyle=dashed,dash=2pt 1pt,linecolor=blue](1.5,1.6){0.6}{115}{160}
		\psarc[linewidth=0.02,linestyle=dashed,dash=2pt 1pt,linecolor=blue](1.5,1.6){0.6}{290}{340}
		\psarc[linewidth=0.02,linestyle=dashed,dash=2pt 1pt,linecolor=red](1.5,1.6){0.6}{200}{250}
		\psline[linewidth=0.02,linestyle=dashed,dash=2pt 1pt,linecolor=red](2.18381,1.78159)(2.4,2.25)
		\psline[linewidth=0.02,linestyle=dashed,dash=2pt 1pt,linecolor=red](2.18381,1.418408)(2.4,0.95)
		\psline[linewidth=0.02,linestyle=dashed,dash=2pt 1pt,linecolor=red](0.1,1.6)(0.7,1.6)
		\psline[linewidth=0.02,linestyle=dashed,dash=2pt 1pt,linecolor=red](1.5,0.2)(1.5,0.8)
		\pscircle[linewidth=0.03,fillstyle=solid,fillcolor=NavyBlue](1.5,2.2){0.25}
		\psline[linewidth=0.03](1.5,2.2)(1.67,2.37)
		\psline[linewidth=0.03](1.5,2.2)(1.33,2.37)
		\psline[linewidth=0.03](1.5,2.2)(1.33,2.03)
		\psline[linewidth=0.03](1.5,2.2)(1.67,2.03)
		\pscircle[linewidth=0.03,fillstyle=solid,fillcolor=lightgray](1.5,1.0){0.20}
		\rput[b]{*0}(1.5,0.92){\scriptsize{$3$}}
		\pscircle[linewidth=0.03,fillstyle=solid,fillcolor=lightgray](0.9,1.6){0.20}
		\pscircle[linewidth=0.03,fillstyle=solid,fillcolor=lightgray](2.1,1.6){0.20}
		\rput[b]{*0}(0.9,1.52){\scriptsize{$3$}}
		\rput[b]{*0}(2.085,1.52){\scriptsize{$4$}}
		\rput[b]{*0}(2.6,0.8){$\pi$}
		\rput[b]{*0}(2.6,2.3){$\pi$}
		\rput[b]{*0}(0.4,1.3){$\pi$}
		\rput[b]{*0}(1.3,0.4){$\pi$}
		\rput[b]{*0}(2.1,2.0){$\sigma$}
		\rput[b]{*0}(0.9,2.0){$\sigma$}
		\rput[t]{*0}(2.1,1.15){$\sigma$}
		\rput[t]{*0}(0.9,1.15){$\pi$}
	\end{pspicture} 
	}} \vspace{-0.6cm}  
	\nonumber\\
	& & \qquad \qquad \qquad \qquad \ - \ \frac{1}{2} \! \! \vcenter{\hbox{
	\begin{pspicture}(3.0,3.2)
		\psarc[linewidth=0.02,linestyle=dashed,dash=2pt 1pt,linecolor=red](1.5,1.6){0.6}{20}{65}
		\psarc[linewidth=0.02,linestyle=dashed,dash=2pt 1pt,linecolor=red](1.5,1.6){0.6}{115}{160}
		\psarc[linewidth=0.02,linestyle=dashed,dash=2pt 1pt,linecolor=red](1.5,1.6){0.6}{290}{340}
		\psarc[linewidth=0.02,linestyle=dashed,dash=2pt 1pt,linecolor=blue](1.5,1.6){0.6}{200}{250}
		\psline[linewidth=0.02,linestyle=dashed,dash=2pt 1pt,linecolor=red](2.18381,1.78159)(2.4,2.25)
		\psline[linewidth=0.02,linestyle=dashed,dash=2pt 1pt,linecolor=red](2.18381,1.418408)(2.4,0.95)
		\psline[linewidth=0.02,linestyle=dashed,dash=2pt 1pt,linecolor=red](0.1,1.6)(0.7,1.6)
		\psline[linewidth=0.02,linestyle=dashed,dash=2pt 1pt,linecolor=red](1.5,0.2)(1.5,0.8)
		\pscircle[linewidth=0.03,fillstyle=solid,fillcolor=RedOrange](1.5,2.2){0.25}
		\psline[linewidth=0.03](1.5,2.2)(1.67,2.37)
		\psline[linewidth=0.03](1.5,2.2)(1.33,2.37)
		\psline[linewidth=0.03](1.5,2.2)(1.33,2.03)
		\psline[linewidth=0.03](1.5,2.2)(1.67,2.03)
		\pscircle[linewidth=0.03,fillstyle=solid,fillcolor=lightgray](1.5,1.0){0.20}
		\rput[b]{*0}(1.5,0.92){\scriptsize{$3$}}
		\pscircle[linewidth=0.03,fillstyle=solid,fillcolor=lightgray](0.9,1.6){0.20}
		\pscircle[linewidth=0.03,fillstyle=solid,fillcolor=lightgray](2.1,1.6){0.20}
		\rput[b]{*0}(0.9,1.52){\scriptsize{$3$}}
		\rput[b]{*0}(2.085,1.52){\scriptsize{$4$}}
		\rput[b]{*0}(2.6,0.8){$\pi$}
		\rput[b]{*0}(2.6,2.3){$\pi$}
		\rput[b]{*0}(0.4,1.3){$\pi$}
		\rput[b]{*0}(1.3,0.4){$\pi$}
		\rput[b]{*0}(2.1,2.0){$\pi$}
		\rput[b]{*0}(0.9,2.0){$\pi$}
		\rput[t]{*0}(2.1,1.15){$\pi$}
		\rput[t]{*0}(0.9,1.15){$\sigma$}
	\end{pspicture}
	}} \! \! \! 
	- \frac{1}{2} \! \! \vcenter{\hbox{
	\begin{pspicture}(3.0,3.2)
		\psarc[linewidth=0.02,linestyle=dashed,dash=2pt 1pt,linecolor=blue](1.5,1.6){0.6}{20}{65}
		\psarc[linewidth=0.02,linestyle=dashed,dash=2pt 1pt,linecolor=blue](1.5,1.6){0.6}{115}{160}
		\psarc[linewidth=0.02,linestyle=dashed,dash=2pt 1pt,linecolor=red](1.5,1.6){0.6}{290}{340}
		\psarc[linewidth=0.02,linestyle=dashed,dash=2pt 1pt,linecolor=blue](1.5,1.6){0.6}{200}{250}
		\psline[linewidth=0.02,linestyle=dashed,dash=2pt 1pt,linecolor=red](0.816188,1.78159)(0.6,2.25)
		\psline[linewidth=0.02,linestyle=dashed,dash=2pt 1pt,linecolor=red](0.816188,1.41841)(0.6,0.95)
		\psline[linewidth=0.02,linestyle=dashed,dash=2pt 1pt,linecolor=red](2.3,1.6)(2.9,1.6)
		\psline[linewidth=0.02,linestyle=dashed,dash=2pt 1pt,linecolor=red](1.5,0.2)(1.5,0.8)
		\pscircle[linewidth=0.03,fillstyle=solid,fillcolor=NavyBlue](1.5,2.2){0.25}
		\psline[linewidth=0.03](1.5,2.2)(1.67,2.37)
		\psline[linewidth=0.03](1.5,2.2)(1.33,2.37)
		\psline[linewidth=0.03](1.5,2.2)(1.33,2.03)
		\psline[linewidth=0.03](1.5,2.2)(1.67,2.03)
		\pscircle[linewidth=0.03,fillstyle=solid,fillcolor=lightgray](1.5,1.0){0.20}
		\rput[b]{*0}(1.5,0.92){\scriptsize{$3$}}
		\pscircle[linewidth=0.03,fillstyle=solid,fillcolor=lightgray](0.9,1.6){0.20}
		\pscircle[linewidth=0.03,fillstyle=solid,fillcolor=lightgray](2.1,1.6){0.20}
		\rput[b]{*0}(0.885,1.52){\scriptsize{$4$}}
		\rput[b]{*0}(2.1,1.52){\scriptsize{$3$}}
		\rput[b]{*0}(0.4,0.8){$\pi$}
		\rput[b]{*0}(0.4,2.3){$\pi$}
		\rput[b]{*0}(2.6,1.3){$\pi$}
		\rput[b]{*0}(1.3,0.4){$\pi$}
		\rput[b]{*0}(2.1,2.0){$\sigma$}
		\rput[b]{*0}(0.9,2.0){$\sigma$}
		\rput[t]{*0}(2.1,1.15){$\pi$}
		\rput[t]{*0}(0.9,1.15){$\sigma$}
	\end{pspicture}
	}} \! \! \!
	- \frac{1}{2} \! \! \vcenter{\hbox{
	\begin{pspicture}(3.0,3.2)
		\psarc[linewidth=0.02,linestyle=dashed,dash=2pt 1pt,linecolor=red](1.5,1.6){0.6}{20}{65}
		\psarc[linewidth=0.02,linestyle=dashed,dash=2pt 1pt,linecolor=red](1.5,1.6){0.6}{115}{160}
		\psarc[linewidth=0.02,linestyle=dashed,dash=2pt 1pt,linecolor=blue](1.5,1.6){0.6}{290}{340}
		\psarc[linewidth=0.02,linestyle=dashed,dash=2pt 1pt,linecolor=red](1.5,1.6){0.6}{200}{250}
		\psline[linewidth=0.02,linestyle=dashed,dash=2pt 1pt,linecolor=red](0.816188,1.78159)(0.6,2.25)
		\psline[linewidth=0.02,linestyle=dashed,dash=2pt 1pt,linecolor=red](0.816188,1.41841)(0.6,0.95)
		\psline[linewidth=0.02,linestyle=dashed,dash=2pt 1pt,linecolor=red](2.3,1.6)(2.9,1.6)
		\psline[linewidth=0.02,linestyle=dashed,dash=2pt 1pt,linecolor=red](1.5,0.2)(1.5,0.8)
		\pscircle[linewidth=0.03,fillstyle=solid,fillcolor=RedOrange](1.5,2.2){0.25}
		\psline[linewidth=0.03](1.5,2.2)(1.67,2.37)
		\psline[linewidth=0.03](1.5,2.2)(1.33,2.37)
		\psline[linewidth=0.03](1.5,2.2)(1.33,2.03)
		\psline[linewidth=0.03](1.5,2.2)(1.67,2.03)
		\pscircle[linewidth=0.03,fillstyle=solid,fillcolor=lightgray](1.5,1.0){0.20}
		\rput[b]{*0}(1.5,0.92){\scriptsize{$3$}}
		\pscircle[linewidth=0.03,fillstyle=solid,fillcolor=lightgray](0.9,1.6){0.20}
		\pscircle[linewidth=0.03,fillstyle=solid,fillcolor=lightgray](2.1,1.6){0.20}
		\rput[b]{*0}(0.885,1.52){\scriptsize{$4$}}
		\rput[b]{*0}(2.1,1.52){\scriptsize{$3$}}
		\rput[b]{*0}(0.4,0.8){$\pi$}
		\rput[b]{*0}(0.4,2.3){$\pi$}
		\rput[b]{*0}(2.6,1.3){$\pi$}
		\rput[b]{*0}(1.3,0.4){$\pi$}
		\rput[b]{*0}(2.1,2.0){$\pi$}
		\rput[b]{*0}(0.9,2.0){$\pi$}
		\rput[t]{*0}(2.1,1.15){$\sigma$}
		\rput[t]{*0}(0.9,1.15){$\pi$}
	\end{pspicture}
	}} \vspace{-0.2cm} \nonumber\\
	& & \qquad \qquad \qquad \qquad \ + \ \frac{1}{2} \! \! \vcenter{\hbox{
	\begin{pspicture}[showgrid=false](3.0,3.2)
		\psarc[linewidth=0.02,linestyle=dashed,dash=2pt 1pt,linecolor=blue](1.5,1.6){0.8}{30}{71}
		\psarc[linewidth=0.02,linestyle=dashed,dash=2pt 1pt,linecolor=blue](1.5,1.6){0.8}{109}{150}
		\psarc[linewidth=0.02,linestyle=dashed,dash=2pt 1pt,linecolor=red](1.5,1.6){0.8}{180}{220}
		\psarc[linewidth=0.02,linestyle=dashed,dash=2pt 1pt,linecolor=blue](1.5,1.6){0.8}{250}{290}
		\psarc[linewidth=0.02,linestyle=dashed,dash=2pt 1pt,linecolor=red](1.5,1.6){0.8}{320}{0}
		\psline[linewidth=0.02,linestyle=dashed,dash=2pt 1pt,linecolor=red](2.35655,1.98865)(2.57274,2.45706)
		\psline[linewidth=0.02,linestyle=dashed,dash=2pt 1pt,linecolor=red](0.643447,1.98865)(0.427259,2.45706)
		\psline[linewidth=0.02,linestyle=dashed,dash=2pt 1pt,linecolor=red](0.957328,0.763086)(0.74114,0.294678)
		\psline[linewidth=0.02,linestyle=dashed,dash=2pt 1pt,linecolor=red](2.04267,0.763086)(2.25886,0.294678)
		\pscircle[linewidth=0.03,fillstyle=solid,fillcolor=NavyBlue](1.5,2.4){0.25}
		\psline[linewidth=0.03](1.5,2.4)(1.67,2.57)
		\psline[linewidth=0.03](1.5,2.4)(1.33,2.57)
		\psline[linewidth=0.03](1.5,2.4)(1.33,2.23)
		\psline[linewidth=0.03](1.5,2.4)(1.67,2.23)
		\pscircle[linewidth=0.03,fillstyle=solid,fillcolor=lightgray](2.27274,1.80706){0.20}
		\rput[b]{*0}(2.27274,1.72706){\scriptsize{$3$}}
		\pscircle[linewidth=0.03,fillstyle=solid,fillcolor=lightgray](0.727259,1.80706){0.20}
		\rput[b]{*0}(0.727259,1.72706){\scriptsize{$3$}}
		\pscircle[linewidth=0.03,fillstyle=solid,fillcolor=lightgray](1.04114,0.944678){0.20}
		\rput[b]{*0}(1.04114,0.864678){\scriptsize{$3$}}
		\pscircle[linewidth=0.03,fillstyle=solid,fillcolor=lightgray](1.95886,0.944678){0.20}
		\rput[b]{*0}(1.95886,0.864678){\scriptsize{$3$}}
		\rput[b]{*0}(2.77274,2.50706){$\pi$}
		\rput[b]{*0}(0.227259,2.50706){$\pi$}
		\rput[b]{*0}(0.54114,0.144678){$\pi$}
		\rput[b]{*0}(2.45886,0.144678){$\pi$}
		\rput[b]{*0}(2.1,2.3){$\sigma$}
		\rput[b]{*0}(0.9,2.3){$\sigma$}
		\rput[b]{*0}(1.5,0.55){$\sigma$}
		\rput[t]{*0}(2.45,1.35){$\pi$}
		\rput[t]{*0}(0.55,1.35){$\pi$}
	\end{pspicture}
	}} \! \! \! 	
	+ \frac{1}{2} \! \! \vcenter{\hbox{
	\begin{pspicture}(3.0,3.2)
		\psarc[linewidth=0.02,linestyle=dashed,dash=2pt 1pt,linecolor=red](1.5,1.6){0.8}{30}{71}
		\psarc[linewidth=0.02,linestyle=dashed,dash=2pt 1pt,linecolor=red](1.5,1.6){0.8}{109}{150}
		\psarc[linewidth=0.02,linestyle=dashed,dash=2pt 1pt,linecolor=blue](1.5,1.6){0.8}{180}{220}
		\psarc[linewidth=0.02,linestyle=dashed,dash=2pt 1pt,linecolor=red](1.5,1.6){0.8}{250}{290}
		\psarc[linewidth=0.02,linestyle=dashed,dash=2pt 1pt,linecolor=blue](1.5,1.6){0.8}{320}{0}
		\psline[linewidth=0.02,linestyle=dashed,dash=2pt 1pt,linecolor=red](2.35655,1.98865)(2.57274,2.45706)
		\psline[linewidth=0.02,linestyle=dashed,dash=2pt 1pt,linecolor=red](0.643447,1.98865)(0.427259,2.45706)
		\psline[linewidth=0.02,linestyle=dashed,dash=2pt 1pt,linecolor=red](0.957328,0.763086)(0.74114,0.294678)
		\psline[linewidth=0.02,linestyle=dashed,dash=2pt 1pt,linecolor=red](2.04267,0.763086)(2.25886,0.294678)
		\pscircle[linewidth=0.03,fillstyle=solid,fillcolor=RedOrange](1.5,2.4){0.25}
		\psline[linewidth=0.03](1.5,2.4)(1.67,2.57)
		\psline[linewidth=0.03](1.5,2.4)(1.33,2.57)
		\psline[linewidth=0.03](1.5,2.4)(1.33,2.23)
		\psline[linewidth=0.03](1.5,2.4)(1.67,2.23)
		\pscircle[linewidth=0.03,fillstyle=solid,fillcolor=lightgray](2.27274,1.80706){0.20}
		\rput[b]{*0}(2.27274,1.72706){\scriptsize{$3$}}
		\pscircle[linewidth=0.03,fillstyle=solid,fillcolor=lightgray](0.727259,1.80706){0.20}
		\rput[b]{*0}(0.727259,1.72706){\scriptsize{$3$}}
		\pscircle[linewidth=0.03,fillstyle=solid,fillcolor=lightgray](1.04114,0.944678){0.20}
		\rput[b]{*0}(1.04114,0.864678){\scriptsize{$3$}}
		\pscircle[linewidth=0.03,fillstyle=solid,fillcolor=lightgray](1.95886,0.944678){0.20}
		\rput[b]{*0}(1.95886,0.864678){\scriptsize{$3$}}
		\rput[b]{*0}(2.77274,2.50706){$\pi$}
		\rput[b]{*0}(0.227259,2.50706){$\pi$}
		\rput[b]{*0}(0.54114,0.144678){$\pi$}
		\rput[b]{*0}(2.45886,0.144678){$\pi$}
		\rput[b]{*0}(2.1,2.3){$\pi$}
		\rput[b]{*0}(0.9,2.3){$\pi$}
		\rput[b]{*0}(1.5,0.55){$\pi$}
		\rput[t]{*0}(2.45,1.35){$\sigma$}
		\rput[t]{*0}(0.55,1.35){$\sigma$}
	\end{pspicture}
	}} \! \! \! 	
	 - \! \! \! \vcenter{\hbox{
	\begin{pspicture}(3.0,3.2)
		\psarc[linewidth=0.02,arrowsize=2pt 3,arrowinset=0]{->}(1.5,1.6){0.8}{30}{60}
		\psarc[linewidth=0.02](1.5,1.6){0.8}{51}{71}
		\psarc[linewidth=0.02,arrowsize=2pt 3,arrowinset=0]{->}(1.5,1.6){0.8}{109}{139}
		\psarc[linewidth=0.02](1.5,1.6){0.8}{130}{150}
		\psarc[linewidth=0.02,arrowsize=2pt 3,arrowinset=0]{->}(1.5,1.6){0.8}{180}{208}
		\psarc[linewidth=0.02](1.5,1.6){0.8}{199}{220}
		\psarc[linewidth=0.02,arrowsize=2pt 3,arrowinset=0]{->}(1.5,1.6){0.8}{250}{277}
		\psarc[linewidth=0.02](1.5,1.6){0.8}{268}{290}
		\psarc[linewidth=0.02,arrowsize=2pt 3,arrowinset=0]{->}(1.5,1.6){0.8}{320}{348}
		\psarc[linewidth=0.02](1.5,1.6){0.8}{339}{0}
		\psline[linewidth=0.02,linestyle=dashed,dash=2pt 1pt,linecolor=red](2.35655,1.98865)(2.57274,2.45706)
		\psline[linewidth=0.02,linestyle=dashed,dash=2pt 1pt,linecolor=red](0.643447,1.98865)(0.427259,2.45706)
		\psline[linewidth=0.02,linestyle=dashed,dash=2pt 1pt,linecolor=red](0.957328,0.763086)(0.74114,0.294678)
		\psline[linewidth=0.02,linestyle=dashed,dash=2pt 1pt,linecolor=red](2.04267,0.763086)(2.25886,0.294678)
		\pscircle[linewidth=0.03,fillstyle=solid,fillcolor=gray](1.5,2.4){0.25}
		\psline[linewidth=0.03](1.5,2.4)(1.67,2.57)
		\psline[linewidth=0.03](1.5,2.4)(1.33,2.57)
		\psline[linewidth=0.03](1.5,2.4)(1.33,2.23)
		\psline[linewidth=0.03](1.5,2.4)(1.67,2.23)
		\pscircle[linewidth=0.03,fillstyle=solid,fillcolor=lightgray](2.27274,1.80706){0.20}
		\rput[b]{*0}(2.27274,1.72706){\scriptsize{$3$}}
		\pscircle[linewidth=0.03,fillstyle=solid,fillcolor=lightgray](0.727259,1.80706){0.20}
		\rput[b]{*0}(0.727259,1.72706){\scriptsize{$3$}}
		\pscircle[linewidth=0.03,fillstyle=solid,fillcolor=lightgray](1.04114,0.944678){0.20}
		\rput[b]{*0}(1.04114,0.864678){\scriptsize{$3$}}
		\pscircle[linewidth=0.03,fillstyle=solid,fillcolor=lightgray](1.95886,0.944678){0.20}
		\rput[b]{*0}(1.95886,0.864678){\scriptsize{$3$}}
		\rput[b]{*0}(2.77274,2.50706){$\pi$}
		\rput[b]{*0}(0.227259,2.50706){$\pi$}
		\rput[b]{*0}(0.54114,0.144678){$\pi$}
		\rput[b]{*0}(2.45886,0.144678){$\pi$}
		\rput[b]{*0}(2.1,2.3){$\psi$}
		\rput[b]{*0}(0.9,2.3){$\psi$}
		\rput[b]{*0}(1.5,0.4){$\psi$}
		\rput[t]{*0}(2.45,1.35){$\psi$}
		\rput[t]{*0}(0.55,1.35){$\psi$}
	\end{pspicture}
	}} \! \! \Biggg), \label{eq:c2} \\[0.2cm]
	\partial_{k}Z_{2,k} & = & \frac{1}{4} \mathcal{V}^{-1} 
	\left.\frac{\mathrm{d}}{\mathrm{d}p^2}\right|_{p^{2}=0}
	\frac{\delta^{4}\partial_{k}\Gamma_{k}}{\delta\pi_{1}(p)
	\delta\pi_{2}(0)\delta\pi_{1}(-p)\delta\pi_{2}(0)} , \\	
	\partial_{k}C_{3,k} & = & \frac{5}{576}\Bigg\lbrace 
	\frac{208}{5}\,\partial_{k}C_{5,k} - 112\, \partial_{k}C_{6,k}
	+ 32\, \partial_{k}C_{7,k} - \frac{224}{5}\, \partial_{k}C_{8,k}
	+ \mathcal{V}^{-1} \bigg(
	\frac{\partial}{\partial p_{1,\mu}}\frac{\partial}{\partial p_{2,\mu}}
	\frac{\partial}{\partial p_{3,\nu}}\frac{\partial}{\partial p_{1,\nu}} \nonumber\\
	& & \qquad\quad - \frac{7}{10}\frac{\partial}{\partial p_{1,\mu}}
	\frac{\partial}{\partial p_{2,\mu}}
	\frac{\partial}{\partial p_{3,\nu}}\frac{\partial}{\partial p_{2,\nu}}
	\bigg)\bigg|_{p_{1}=p_{2}=p_{3}=0}
	\frac{\delta^{4}\partial_{k}\Gamma_{k}}{\delta\pi_{1}(p_{1})
	\delta\pi_{2}(p_{2})\delta\pi_{1}(p_{3})\delta\pi_{2}(-p_{1}-p_{2}-p_{3})}
	\Bigg\rbrace , \label{eq:c3} \\
	\partial_{k}C_{4,k} & = & - \frac{1}{288}\Bigg\lbrace 
	- 16\, \partial_{k}C_{5,k} - 400\, \partial_{k}C_{6,k}
	+ 32\, \partial_{k}C_{7,k} - 160\, \partial_{k}C_{8,k}
	+ \mathcal{V}^{-1} \bigg(
	\frac{\partial}{\partial p_{1,\mu}}\frac{\partial}{\partial p_{2,\mu}}
	\frac{\partial}{\partial p_{3,\nu}}\frac{\partial}{\partial p_{1,\nu}} \nonumber\\
	& & \qquad\quad\ - \frac{5}{2}\frac{\partial}{\partial p_{1,\mu}}
	\frac{\partial}{\partial p_{2,\mu}}
	\frac{\partial}{\partial p_{3,\nu}}\frac{\partial}{\partial p_{2,\nu}}
	\bigg)\bigg|_{p_{1}=p_{2}=p_{3}=0}
	\frac{\delta^{4}\partial_{k}\Gamma_{k}}{\delta\pi_{1}(p_{1})
	\delta\pi_{2}(p_{2})\delta\pi_{1}(p_{3})\delta\pi_{2}(-p_{1}-p_{2}-p_{3})}
	\Bigg\rbrace , \\
	\partial_{k}C_{5,k} & = & \frac{1}{96} \mathcal{V}^{-1} 
	\bigg(\frac{\partial}{\partial p_{2,\mu}}\frac{\partial}{\partial p_{2,\mu}}
	\frac{\partial}{\partial p_{2,\nu}}\frac{\partial}{\partial p_{3,\nu}} \nonumber\\
	& & \qquad\qquad\quad - \frac{1}{2}\frac{\partial}{\partial p_{2,\mu}}
	\frac{\partial}{\partial p_{2,\mu}}
	\frac{\partial}{\partial p_{2,\nu}}\frac{\partial}{\partial p_{2,\nu}}
	\bigg)\bigg|_{p_{2}=p_{3}=0}
	\frac{\delta^{4}\partial_{k}\Gamma_{k}}{\delta\pi_{1}(-p_{2}-p_{3})
	\delta\pi_{2}(p_{2})\delta\pi_{1}(p_{3})\delta\pi_{2}(0)}, \\
	\partial_{k}C_{6,k} & = & - \frac{1}{96}\Bigg\lbrace 
	160\, \partial_{k}C_{5,k} + 64\, \partial_{k}C_{7,k} + \mathcal{V}^{-1}
	\frac{\partial}{\partial p_{2,\mu}}\frac{\partial}{\partial p_{4,\mu}}
	\frac{\partial}{\partial p_{2,\nu}}\frac{\partial}{\partial p_{4,\nu}}
	\bigg|_{p_{2}=p_{4}=0}
	\frac{\delta^{4}\partial_{k}\Gamma_{k}}{\delta\pi_{1}(-p_{2}-p_{4})
	\delta\pi_{2}(p_{2})\delta\pi_{1}(0)\delta\pi_{2}(p_{4})} \nonumber\\
	& & \qquad\quad\, - \frac{1}{12} \mathcal{V}^{-1}
	\frac{\partial}{\partial p_{\mu}}
	\frac{\partial}{\partial p_{\mu}}
	\frac{\partial}{\partial p_{\nu}}\frac{\partial}{\partial p_{\nu}}
	\bigg|_{p=0}
	\frac{\delta^{4}\partial_{k}\Gamma_{k}}{\delta\pi_{1}(-p)
	\delta\pi_{2}(0)\delta\pi_{1}(p)\delta\pi_{2}(0)}
	\Bigg\rbrace , \\
	\partial_{k}C_{7,k} & = & - \frac{1}{384} \mathcal{V}^{-1}
	\frac{\partial}{\partial p_{\mu}}\frac{\partial}{\partial p_{\mu}}
	\frac{\partial}{\partial p_{\nu}}\frac{\partial}{\partial p_{\nu}}
	\bigg|_{p=0}
	\frac{\delta^{4}\partial_{k}\Gamma_{k}}{\delta\pi_{1}(-p)
	\delta\pi_{2}(p)\delta\pi_{1}(0)\delta\pi_{2}(0)}, \\
	\partial_{k}C_{8,k} & = & \frac{1}{96}\Bigg\lbrace 
	160\, \partial_{k}C_{5,k} + 64\, \partial_{k}C_{7,k} + \mathcal{V}^{-1}
	\frac{\partial}{\partial p_{2,\mu}}\frac{\partial}{\partial p_{4,\mu}}
	\frac{\partial}{\partial p_{2,\nu}}\frac{\partial}{\partial p_{4,\nu}}
	\bigg|_{p_{2}=p_{4}=0}
	\frac{\delta^{4}\partial_{k}\Gamma_{k}}{\delta\pi_{1}(-p_{2}-p_{4})
	\delta\pi_{2}(p_{2})\delta\pi_{1}(0)\delta\pi_{2}(p_{4})} \nonumber\\
	& & \qquad\ - \frac{5}{24} \mathcal{V}^{-1}
	\frac{\partial}{\partial p_{\mu}}
	\frac{\partial}{\partial p_{\mu}}
	\frac{\partial}{\partial p_{\nu}}\frac{\partial}{\partial p_{\nu}}
	\bigg|_{p=0}
	\frac{\delta^{4}\partial_{k}\Gamma_{k}}{\delta\pi_{1}(-p)
	\delta\pi_{2}(0)\delta\pi_{1}(p)\delta\pi_{2}(0)}
	\Bigg\rbrace \label{eq:c8} .
\end{IEEEeqnarray}
As before, $p$ or $p_{i}$ with $i \in \lbrace 1,2,3,4 \rbrace$ represents the external momentum of
the corresponding leg. The diagrams shown on the right side of the equations include all 
possible permutations of the external fields. Five- and six-point vertices are
truncated.

Note that the wave-function renormalizations $Z_{k}^{\sigma}$ and $Z_{k}^{\pi}$ 
defined in Eqs.\ (\ref{eq:Zsigma}) and (\ref{eq:Zpi}) are evaluated at 
vanishing external momentum, $p^{2} = 0$, i.e.,\vspace{3pt}
\begin{IEEEeqnarray}{rCl}
	Z_{k}^{\sigma} & = & Z_{k} + 2\sigma^{2}\left(Z_{2,k} + C_{2,k}\right),\\
	Z_{k}^{\pi} & = & Z_{k} + 2\sigma^{2} Z_{2,k}.\vspace{3pt}
\end{IEEEeqnarray}
Due to the full $O(4)$-symmetric truncation in Eq.\ (\ref{eq:truncation}), the
equations of $Z_{k}^{\sigma}$ and $Z_{k}^{\pi}$ also include all diagrams with momentum-dependent
four-point vertices. This is in contrast to Ref.\ \cite{Eser:2018jqo}, where we only considered
momentum-dependent four-pion interactions.

To derive the analytical flow equations we made use of the \texttt{Mathematica} 
packages \texttt{FeynCalc} \cite{Mertig:1990an, Shtabovenko:2016sxi}, \texttt{DoFun} 
\cite{Huber:2011qr}, and \texttt{FormTracer} \cite{Cyrol:2016zqb}.
\end{widetext}

\section{Solving the flow equations}
\label{sec:solv_floweqns}

We solve the system of coupled flow equations using a Taylor-polynomial ansatz 
for the effective potential,
\begin{equation}
	U_{k}\left(\rho\right) = \sum_{n=1}^{N} 
	\frac{\alpha_{n,k}}{n!} \left(\rho - \chi\right)^{n}, \label{eq:taylor}
\end{equation}
where the expansion point $\chi$ is scale independent. This choice of $\chi$ follows
the considerations of Ref.\ \cite{Pawlowski:2014zaa}. The ansatz (\ref{eq:taylor}) leads 
to stable results for $N = 6$, in the sense that the Taylor solution coincides with the results from
a grid discretization.

The flow equations for the Taylor coefficients $\alpha_{n,k}$ are obtained from 
derivatives of the flow equation of the effective potential,
\begin{equation}
	\partial_{k}\alpha_{n,k} = \left.\frac{\partial^{n}}
	{\partial\rho^{n}}\right\vert_{\chi}
	\partial_{k} U_{k} .
\end{equation}
\begin{table}[b!]
	\caption{\label{tab:UVpars}UV parameters ($\Lambda = 500$ MeV).}
	\begin{ruledtabular}
		\begin{tabular}{ccccc}
		$\chi$ & $\alpha_{1,\Lambda}$ & $\alpha_{2,\Lambda}$ & $h_{\mathrm{ESB}}$ & $y_{\Lambda}$ \\ 
		\colrule
		$61.34^{2}\ \mathrm{MeV}^{2}$ & $600^{2}\ \mathrm{MeV}^{2}$ & $1.4$ & 
		$2.75 \times 10^{6}\ \mathrm{MeV}^{3}$ & $8.96$
		\end{tabular}
	\end{ruledtabular}
\end{table}
The equations for the wave-function renormalization factors, the Yukawa coupling, 
and the higher-derivative couplings are also evaluated at the constant expansion 
point $\chi$, which is chosen to be slightly larger than the IR minimum 
$\rho_{0}$ of the potential.
\begin{figure}[t!]
	\centering
		\includegraphics{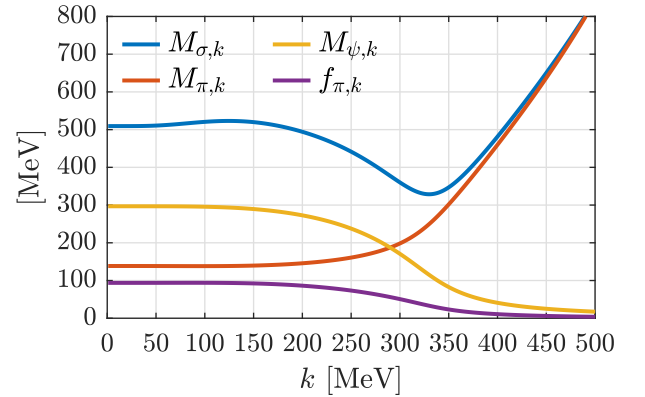}
	\caption{Scale evolution of the renormalized sigma, pion, and quark 
	masses, as well as the pion decay constant; $k_{\mathrm{IR}} = 1\ \mathrm{MeV}$.}
	\label{fig:masses}
\end{figure}

We initialize the set of flow equations at the UV cutoff $\Lambda = 500\ \mathrm{MeV}$, which 
is a typical scale for the QMM and other NJL-like models. In order to capture quark dynamics, 
this scale somewhat exceeds the ones suggested by recent investigations \cite{Braun:2014ata, 
Mitter:2014wpa, Cyrol:2017ewj}, which have shown that the range of validity of these models is 
around 300 MeV.

At the cutoff $\Lambda$, the effective potential is tuned such that we reproduce
experimental data for the meson and quark masses as well as for the pion decay
constant in the IR, cf.\ Ref.\ \cite{Tanabashi:2018oca}. The final parameters of this 
tuning process, including the ESB parameter $h_{\mathrm{ESB}}$ and the Yukawa coupling
$y_{\Lambda}$, are quoted in Table \ref{tab:UVpars}.
The coefficients $\alpha_{n,\Lambda}$, which are not shown, are set to zero, 
i.e., $\alpha_{n,\Lambda} = 0$ for $2 < n \leq 6$.

The wave-function renormalization factors are initialized with a value of 
one and the higher-derivative couplings start at zero in the UV. This means that
the couplings $\tilde{C}_{2,k}$ and $\tilde{Z}_{2,k}$, as well as $\tilde{C}_{i,k}$, $i=3,\ldots,8$,
are only dynamically generated during the integration from the UV to the IR.
\begin{figure}[t!]
	\centering
		\includegraphics{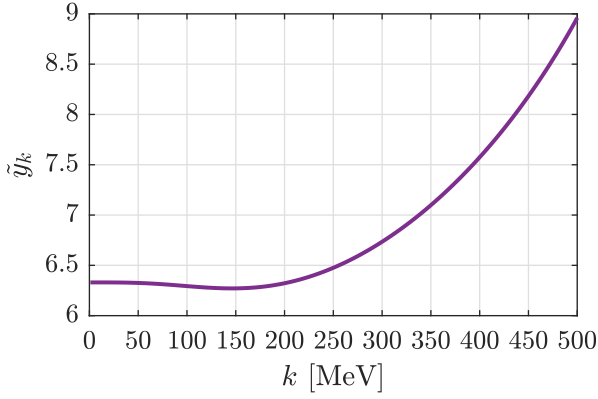}
	\caption{Scale evolution of the renormalized Yukawa coupling 
	$\tilde{y}_{k}$; $k_{\mathrm{IR}} = 1\ \mathrm{MeV}$.}
	\label{fig:yc}
\end{figure}

The scale-dependent ``bare'' masses of the $\sigma$, $\pi$, and quark fields are computed 
from the effective potential. They read
\begin{IEEEeqnarray}{rCl}
	m_{\sigma,k}^{2} & = & 2 U_{k}'(\rho_{0,k}) + 
	4\rho_{0,k} U_{k}''(\rho_{0,k}) , \label{eq:msigma} \\
	m_{\pi,k}^{2} & = & 2 U_{k}'(\rho_{0,k}) , \label{eq:mpi} \\
	m_{\psi,k}^{2} & = & \frac{y_{k}^{2}}{4}\rho_{0,k} , \label{eq:mquark}
\end{IEEEeqnarray}
where $\rho_{0,k}$ denotes the scale-dependent minimum of the potential
\begin{equation}
	U_{k}\left(\rho\right) - h_{\mathrm{ESB}}\sqrt{\rho} .
\end{equation}
It is related to the pion decay constant via the PCAC relation \cite{Eser:2018jqo},
\begin{equation}
	f_{\pi,k} = \sqrt{Z_{k}^{\pi} \rho_{0,k}}
	\equiv \sqrt{Z_{k}^{\pi}} \sigma_{0,k}, \label{eq:PCAC}
\end{equation}
where $\sigma_{0,k}$ denotes the scale-dependent vacuum expectation
value of the $\sigma$ field. The renormalized masses corresponding to Eqs.\ (\ref{eq:msigma}), 
(\ref{eq:mpi}), and (\ref{eq:mquark}) are given by
\begin{equation}
	M_{\sigma,k}^{2} = \frac{m_{\sigma,k}^{2}}
	{Z_{k}^{\sigma}} ,\quad
	M_{\pi,k}^{2} = \frac{m_{\pi,k}^{2}}
	{Z_{k}^{\pi}} ,\quad
	M_{\psi,k}^{2} = \frac{m_{\psi,k}^{2}}
	{({Z_{k}^{\psi}})^{2}} .
\end{equation}

After the initialization of the flow, the system of equations is integrated down to 
$1\ \mathrm{MeV}$, which is a valid IR cutoff less than one percent of $\Lambda$, at 
which all $k$-dependent quantities are entirely frozen out.

Figure \ref{fig:masses} shows the scale dependence of the renormalized masses
and the pion decay constant, which is related to the expectation value of the
sigma field via Eq.\ (\ref{eq:PCAC}). The latter serves as an (approximate) order parameter 
for the dynamical $O(4)$-symmetry breaking [the $O(4)$ invariance of the model is also 
explicitly broken by the parameter $h_{\mathrm{ESB}} \neq 0$].

In the regime close to the UV, the mesonic fields are approximately degenerate in
mass. This is required by the almost restored $O(4)$ symmetry, as indicated by the small
value of the expectation value of the sigma field. The quarks are comparatively
light in the same regime and they therefore dominate the RG flow.

For lower energies, the QMM undergoes a crossover transition and, hence, 
the spontaneous breaking of the $O(4)$ symmetry becomes increasingly obvious. At $k \lesssim 50\ \mathrm{MeV}$, 
the renormalized masses as well as the pion decay constant approach their ``physical'' IR-limit values:
$M_{\sigma,k_{\mathrm{IR}}} = 509.4\ \mathrm{MeV}$, $M_{\pi,k_{\mathrm{IR}}} \equiv 
\mathcal{M}_{\Pi,k_{\mathrm{IR}}} = 138.5\ \mathrm{MeV}$, 
$M_{\psi,k_{\mathrm{IR}}} = 296.8\ \mathrm{MeV}$, and $f_{\pi,k_{\mathrm{IR}}} \equiv f_{\pi} 
= 93.8\ \mathrm{MeV}$.
\begin{figure}[t!]
	\centering
		\includegraphics{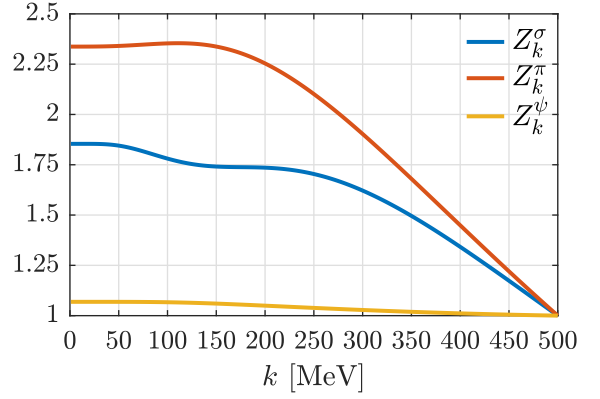}
	\caption{Scale evolution of the bosonic and fermionic wave-function 
	renormalization factors $Z_{k}^{\sigma}$, $Z_{k}^{\pi}$, and $Z_{k}^{\psi}$; 
	$k_{\mathrm{IR}} = 1\ \mathrm{MeV}$.}
	\label{fig:wave}
\end{figure}

In Fig.\ \ref{fig:yc} we present the RG flow of the renormalized Yukawa coupling
$\tilde{y}_{k}$. Starting with a value of $8.96$ at the UV scale, it shrinks during the 
integration and picks up a value of $6.33$ in the IR.

The bosonic and fermionic wave-function renormalization factors, $Z_{k}^{\sigma}$, $Z_{k}^{\pi}$, and 
$Z_{k}^{\psi}$, respectively, are plotted in Fig.\ \ref{fig:wave}. The numerical values 
of $Z_{k}^{\pi}$ and $Z_{k}^{\psi}$ give the correction from unrenormalized to renormalized 
quantities in Eq.\ (\ref{eq:rquantities}), e.g., the higher-derivatve couplings
$\tilde{C}_{2,k}$, $\tilde{Z}_{2,k}$, and $\tilde{C}_{i,k}$, $i=3,\ldots,8$,
differ from $C_{2,k}$, $Z_{2,k}$, and $C_{i,k}$, $i=3,\ldots,8$, by a factor of 
$1/(Z_{k}^{\pi})^{2} \simeq 0.18$ in the IR; $Z_{k_{\mathrm{IR}}}^{\pi} = 2.34$. 
In contrast to the bosonic fields, the quark-field renormalization remains moderate, 
$Z_{k}^{\psi} = 1.07$ at $k = 1\ \mathrm{MeV}$.
The factor $Z_{k}^{\sigma}$, see also Fig.\ \ref{fig:wave}, 
has an IR value of $1.85$. The bosonic wave-function renormalization factors 
are evaluated at the constant expansion point $\chi \equiv \sigma^{2}$, as introduced in the
Taylor polynomial in Eq.\ (\ref{eq:taylor}), and, therefore, necessarily differ over 
the entire $k$ range.

\bibliography{bib_LECs}

\end{document}